\documentclass[times]{aastex62}
\usepackage{graphicx}
\usepackage{dcolumn, color}
\usepackage{bm}
\usepackage[fleqn]{amsmath}
\usepackage{amsfonts,amssymb} 
\usepackage[e]{esvect}
\usepackage{hyperref}
\usepackage{bigints}
\usepackage{xparse}
\usepackage{rotating}
\usepackage[utf8]{inputenc}
\usepackage[T1]{fontenc}
\usepackage{mathptmx}
\usepackage{subfigure}
\renewcommand{\vv}[1]{\ensuremath{\vec{#1}}}

\newcommand{\erf}{\ensuremath{\mathrm{erf}}}

\newcommand{\tr}[1]{\ensuremath{\mathrm{tr}(\tensor{#1})}}
\NewDocumentCommand\hyp{oo}{
  \IfNoValueTF{#1}{\ensuremath{\ \mathrm{_{[]}F_{[]}}}}
     {\IfNoValueTF{#2}{\mbox{\ensuremath{\ _{[#1]}\mathrm{F}_{[]}}}}
       {\mbox{\ensuremath{\ _{[#1]}\mathrm{F}_{[#2]}}}}
}}

\NewDocumentCommand\Ika{oo}{
  \IfNoValueTF{#1}{\ensuremath{\ {\matcal{U}_{[#2]}}}}
     {\IfNoValueTF{#2}{\mbox{\ensuremath{\ _{[#1]}\mathcal{U}_{[]}}}}
       {\mbox{\ensuremath{\ _{[#1]}\mathcal{U}_{[#2]}}}}
     }
   }
\NewDocumentCommand\ika{oo}{
  \IfNoValueTF{#1}{\ensuremath{\ {\mathrm{_{[]}u}}}}
     {\IfNoValueTF{#2}{\mbox{\ensuremath{\ _{[#1]}\mathrm{u}}}}
       {\mbox{\ensuremath{\ _{[#1]}\mathrm{u}}}}
}}
\NewDocumentCommand\Jka{oo}{
  \IfNoValueTF{#1}{\mbox{\ensuremath{\ \ _{{[]}}\mathcal{W}_{[]}}}}
     {\IfNoValueTF{#2}{\mbox{\ensuremath{\ _{[#1]}\mathcal{W}_{[]}}}}
       {\mbox{\ensuremath{\ _{[#1]}\mathcal{W}_{[#2]}}}}
}}
\NewDocumentCommand\jka{oo}{
  \IfNoValueTF{#1}{\ensuremath{\ \mathrm{_{[]}w}}}
     {\IfNoValueTF{#2}{\mbox{\ensuremath{\ _{[#1]}\mathrm{w}}}}
       {\mbox{\ensuremath{\ _{[#1]}\mathrm{w}}}}
}}
\NewDocumentCommand\Vka{ooo}{
  \IfNoValueTF{#1}{\mbox{\ensuremath{\ \ _{{[]}}\mathcal{V}_{[]}^{#3}}}}
     {\IfNoValueTF{#2}{\mbox{\ensuremath{\ _{[#1]}\mathcal{V}_{[]}^{#3}}}}
       {\mbox{\ensuremath{\ _{[#1]}\mathcal{V}_{[#2]}^{#3}}}}
}}\NewDocumentCommand\sinf{oo}{
  \IfNoValueTF{#1}{\ensuremath{\sin\varphi}}
     {\IfNoValueTF{#2}{\ensuremath{\sin^{#1}\varphi}}
       {\ensuremath{\sin^{#1}#2}}
}}
\NewDocumentCommand\cosf{oo}{
  \IfNoValueTF{#1}{\ensuremath{\cos\varphi}}
     {\IfNoValueTF{#2}{\ensuremath{\cos^{#1}\varphi}}
       {\ensuremath{\cos^{#1}#2}}
}}
\NewDocumentCommand\sint{oo}{
  \IfNoValueTF{#1}{\ensuremath{\sin\vartheta}}
     {\IfNoValueTF{#2}{\ensuremath{\sin^{#1}\vartheta}}
       {\ensuremath{\sin^{#1}#2}}
}}
\NewDocumentCommand\cost{oo}{
  \IfNoValueTF{#1}{\ensuremath{\cos\vartheta}}
     {\IfNoValueTF{#2}{\ensuremath{\cos^{#1}\vartheta}}
       {\ensuremath{\cos^{#1}#2}}
}}
\newcommand{\icyl}{\ensuremath{
    \int\limits_{0}^{{2\pi}}\int\limits_{0}^{\infty}
    \int\limits_{-\infty}^{\infty}}}
\newcommand{\isph}{\ensuremath{\int\limits_{0}^{\infty}\int\limits_{0}^{2\pi} \int\limits_{0}^{\pi}}}
\newcommand{\Tp}{\ensuremath{\Theta\pa}}
\newcommand{\Ts}{\ensuremath{\Theta\se}}
\newcommand{\as}{\ensuremath{\alpha\se}}
\renewcommand{\ap}{\ensuremath{\alpha\pa}}

\newcommand{\dw}{\ensuremath{\mathrm d}w}
\newcommand{\dx}{\ensuremath{\mathrm d}x}

\newcommand{\dr}{\ensuremath{\mathrm d}r}
\newcommand{\dt}{\ensuremath{\mathrm dt}}
\newcommand{\dtheta}{\ensuremath{\mathrm d}\vartheta}
\newcommand{\dphi}{\ensuremath{\mathrm d}\varphi}
\newcommand{\pr}{\ensuremath{_{\mathrm{RPBK}}}}
\newcommand{\su}{\ensuremath{_{\mathrm{RBK}}}}
\newcommand{\pa}{\ensuremath{_{\parallel}}}
\newcommand{\se}{\ensuremath{_{\perp}}}
\newcommand{\vols}{\ensuremath{w^{2} \sint \mathrm{d}\vartheta\mathrm{d}\varphi\mathrm{d}w}}
\newcommand{\vls}{\ensuremath{ \sint \mathrm{d}\vartheta\mathrm{d}\varphi\mathrm{d}w}}

\newcommand{\volc}{\ensuremath{w\se\mathrm{d}w\pa\mathrm{d}w\se\mathrm{d}\vartheta}}
\newcommand{\voc}{\ensuremath{w_{\perp}\mathrm{d}w\pa\mathrm{d}w\se}}
\newcommand{\vo}{\ensuremath{\mathrm{d}w\pa\mathrm{d}w\se\mathrm{d}\vartheta}}
\providecommand\tensor[1]{\ensuremath{\overleftrightarrow{#1}}}

\newcommand{\GA}[1]{\ensuremath{\Gamma\left(#1\right)}}
\newcommand{\kB}{\ensuremath{k_{\mathrm{B}}}}

\providecommand\aap{Astron. \& Astrophys.}
\providecommand\mnras{Mon. Not. R. Astron. Soc. }
\providecommand\apss{Astrophys. Space Sci.}
\providecommand\apj{Astrophs. J.}
\let\oldsqrt\sqrt
\def\sqrt{\mathpalette\DHLhksqrt}
\def\DHLhksqrt#1#2{%
\setbox0=\hbox{$#1\oldsqrt{#2\,}$}\dimen0=\ht0
\advance\dimen0-0.2\ht0
\setbox2=\hbox{\vrule height\ht0 depth -\dimen0}%
{\box0\lower0.4pt\box2}}

\allowdisplaybreaks

\received{}
\revised{}
\accepted{}
\submitjournal{ApJS}
\shorttitle{Anisotropic regularized $\kappa$-distributions}
\shortauthors{Scherer et al.}
\begin{document}
\title{Moments of the anisotropic regularized $\kappa$-distributions}
\correspondingauthor{K. Scherer}
\email{kls@tp4-rub.de}

\author[0000-0002-9530-1396]{Klaus Scherer}
\affiliation{Institut f\"ur Theoretische Physik, Lehrstuhl IV:
  Plasma-Astroteilchenphysik, Ruhr-Universit\"at Bochum, D-44780 Bochum,
  Germany}
\affiliation{Research Department, Plasmas with Complex Interactions,
  Ruhr-Universit\"at Bochum, 44780 Bochum, Germany}

\author[0000-0002-8508-5466]{Marian Lazar}
\affiliation{Institut f\"ur Theoretische Physik, Lehrstuhl IV:
  Plasma-Astroteilchenphysik, Ruhr-Universit\"at Bochum, D-44780 Bochum,
  Germany}
\affiliation{Centre for Mathematical Plasma Astrophysics,
  Celestijnenlaan 200B, 3001 Leuven Belgium }

\author{Edin Husidic}
\affiliation{Institut f\"ur Theoretische Physik, Lehrstuhl IV:
  Plasma-Astroteilchenphysik, Ruhr-Universit\"at Bochum, D-44780 Bochum,
  Germany}

\author[0000-0002-9151-5127]{Horst Fichtner}
\affiliation{Institut f\"ur Theoretische Physik, Lehrstuhl IV:
  Plasma-Astroteilchenphysik, Ruhr-Universit\"at Bochum, D-44780 Bochum,
  Germany}
\affiliation{Research Department, Plasmas with Complex Interactions,
  Ruhr-Universit\"at Bochum, 44780 Bochum, Germany} \date{\today}

\begin{abstract}
  For collisionless (or collision-poor) plasma populations which are
  well described by the $\kappa$-distribution
  functions (also known as the Kappa or Lorentzian power-laws) a
  macroscopic interpretation has remained largely questionable,
  especially because of the diverging moments of these
  distributions. Recently significant progress has been made by
  introducing a generic regularization for the isotropic
  $\kappa$-distribution,
  which resolves this critical limitation. Regularization is here
  applied to the anisotropic forms of $\kappa$-distributions,
  commonly used to describe temperature anisotropies, and skewed or
  drifting distributions of beam-plasma systems. These regularized
  distributions admit non-diverging moments which are provided for all
  positive $\kappa$,
  opening promising perspectives for a macroscopic (fluid-like)
  characterization of non-ideal plasmas.
\end{abstract}

\section{\label{sec:1}Introduction}

Macroscopic models of plasmas systems are constructed on the principal
(zeroth to second order) moments of velocity distributions of their
particles. In collisionless or collision-poor plasmas, ubiquitous in
space or fusion setups, the velocity distributions of charged
particles are far from thermal equilibrium, exhibiting non-Maxwellian
features like temperature anisotropies, beaming (or drifting)
components and suprathermal tails. Despite these evidences theoretical
predictions are still largely based on idealized scenarios assuming
particles well described by bi-Maxwellian distributions, allowing for a
straightforward definition of the macroscopic parameters using the
main moments of the distribution.  Introduced 50 years ago
\cite{Olbert-1968, Vasyliunas-1968}, as a generalization of the
idealized Maxwellian, $\kappa$-distributions
have gained much notoriety in the last decades, especially for their
ability to reproduce the velocity and energy distributions of plasma
particles in the solar wind and planetary magnetospheres, see the
review by \citet{Pierrard-Lazar-2010}. Suprathermal populations
present in these environments enhance the high-energy tails of the
observed distributions which are successfully described by the
$\kappa$-distribution
functions. These power-laws have been widely invoked to study various
kinetic effects in non-ideal plasmas, e.g., generation of suprathermal
particle populations, solar wind acceleration, particle heating, waves
dissipation or instabilities as local sources of electromagnetic
fluctuations. The existence of $\kappa$-distributions
in space plasmas is more than obvious, but their relevance has become
questionable owing to their limitations in conveying a macroscopic
approach, with non-diverging moments $M_l$
restricted to low orders $l < 2 \kappa - 1$.
\citet{Scherer-etal-2017} have recently introduced a regularization of
$\kappa$-distributions
which can resolve this limitation by removing all singularities from
the theory. Introduced for simple isotropic distributions, the
regularized $\kappa$-distribution (RKD)
\begin{align}
f_{R}(\kappa, w) &=  n_0 N_{R} \left(1 + \frac{w^{2}}{\kappa}\right)^{-\kappa-1}
e^{-\alpha^{2} w^{2}} \label{e1}
\end{align}
combines the standard (isotropic) $\kappa$-distribution 
\begin{align}
f_{K}(\kappa, w) &=  n_0 N_{K} \left(1 + \frac{w^{2}}{\kappa}\right)^{-\kappa-1} \label{e2}
\end{align}
with a Maxwellian cutoff $\exp (-\alpha^{2} w^{2})$,
where the regularization parameter  $0 < \alpha < 1$
should be small enough to conform with the observations and
theoretical predictions.  In these expressions $w = v / \Theta$
denotes particle velocity, usually normalized to a convenient speed
$\Theta$,
$n_0$
is the number density of particles, and $N_R$
and $N_K$
are normalization constants, with $N_R$
defined in \citet{Scherer-etal-2017}, Eq.~(9), and
$N_K =
{\Gamma(\kappa)}/{\Gamma\left(\kappa-\frac{1}{2}\right)}/(\sqrt{\pi^{3}\kappa}\Theta^{3})$.
By contrast to the standard $\kappa$-distribution,
all velocity moments of the RKD are convergent for any positive
$\kappa$,
and have been expressed analytically in \citet{Scherer-etal-2017}.

In the present paper we consider more complex distribution functions
capable to reproduce kinetic anisotropies of plasma particles in
collisionless plasmas from space, like, anisotropic temperatures,
e.g.,
$A \equiv T_\perp / T_\parallel
\ne 1$, usually defined with respect to the direction of an ambient
magnetic field, or/and field aligned beaming (or drifting)
components. The anisotropic $\kappa$-models
can reproduce the gyrotropic distributions of suprathermal (halo)
populations measured in general in the solar wind, e.g, or bi-Kappa
distribution functions
\cite{Maksimovic-etal-2005, Stverak-etal-2008}, or during energetic
events, like fast winds or coronal mass ejections, when the
suprathermal tails of the observed distributions become skewed in the
presence of field-aligned counter-moving beams (or double Strahls) and
resemble a product-bi-Kappa distribution, see \citet{Lazar-etal-2012}.
These anisotropic distribution functions are tabulated by
\citet[][see their Table~1]{Summers-Thorne-1991} and are employed to
explain the observed fluctuations, generated spontaneously
\cite{Vinas-etal-2015, Kim-etal-2017} or stimulated by various wave
instabilities, e.g., firehose
\cite{Astfalk-Jenko-2016, Lazar-etal-2017a}, cyclotron
\cite{Lazar-etal-2011, Lazar-Poedts-2014, Lazar-etal-2015,
  Eliasson-Lazar-2015, Lazar-etal-2016, dosSantos-etal-2017,
  Ziebell-Gaelzer-2017, Lazar-etal-2018},
or mirror instability
\cite{Leubner-Schupfer-2002, Shaaban-etal-2018}.

The regularized forms of anisotropic $\kappa$-distribution
functions are introduced in section~\ref{sec:2}, and then, in
section~\ref{sec:3} we evaluate the principal moments (zero to third
order) of these distributions. The moments are calculated for
distribution functions reproducing temperature anisotropies in the
rest frame as well as in a drifting reference frame, in order to
include anisotropic drifting components (beams, Strahls, counterbeams,
etc.).  Potential applications are discussed in section~\ref{sec:4}
and conclusions are formulated in section~\ref{sec:5}. Necessary
details from derivations, including mathematical definitions and
symbols used, are given in the appendices. Appendix~\ref{appVec}
contains a short note on the vector calculus, while some details of
the moment calculations are shown in Appendix~\ref{appMom}. The
evaluation of the integrals is presented in Appendix~\ref{appInt}, and
finally the special treatment of the most probable speeds and heat
flows is discussed in Appendix~\ref{appMost}.

\section{Regularized anisotropic {\large$\kappa$}-distributions\label{sec:2} }

Using the same technique as in~\citet{Scherer-etal-2017}, here we
regularize the anisotropic $\kappa$-distributions
which are often used in space physics to model anisotropic
temperature.

\begin{align}\label{eq:mod1}\nonumber
f\su&(\kappa,\alpha\pa,\alpha\se,w\pa,w\se)
  = n_{0}N\su\\
      &\left(1+ \frac{w\pa^{2}}{\kappa} +
      \frac{w\se^{2}}{\kappa}\right)^{-\kappa-1}
      \mathrm{e}^{-\alpha\pa^{2}w\pa^{2}-\alpha\se^{2}w\se^{2}},
\end{align} 
is the regularized bi-$\kappa$ (RBK) distribution function, and 
\begin{align}\label{eq:mod}
 f\pr (\kappa\pa,\kappa\se,\alpha\pa,\alpha\se,w\pa,w\se,s\pa.s\se) = n_{0}N\pr
    \left(1+ \frac{w\pa^{2}}{\kappa\pa}\right)^{-\kappa\pa-s\pa}
   \left(1+ \frac{w\se^{2}}{\kappa\se}\right)^{-\kappa\se-s\se}
   \mathrm{e}^{-\alpha\pa^{2}w\pa^{2}-\alpha\se^{2}w\se^{2}}
\end{align}
is the regularized product-bi-$\kappa$
(RPK) distribution, with dimensionless velocities $w\pa,w\se$
(for $w$ see also above)
\begin{align*}\label{eq:rpbk}
  w &= \frac{v}{\Theta}, \qquad  &w\pa &= \frac{v\pa}{\Theta\pa},\\
   w\se &= \frac{v\se}{\Theta\se}  \qquad &\vv{w}&=w\pa\vv{e}\pa+\vv{w}\se
\end{align*}
where $\vec{w}\se$
is a vector in a plane perpendicular to $\vec{e}\pa$.
We adopt the general case with two distinct positive cut-off
parameters $\ap\ne\as$.
In the limit of $\alpha\pa, \alpha\se \to 0$
$f\su$
reduces to the standard bi-$\kappa$
(BK) also known as bi-Kappa or bi-Lorentzian distribution function,
and $f\pr$
reduces to the standard product-bi-$\kappa$
(PBK), for $s\pa=1$
and $s\se=1$.
Both these two standard forms are largely invoked in studies of
anisotropic temperatures and their implications (see the
introduction).  Notice that with $f\pr$
we have the ability to describe decoupled parallel and perpendicular
components, with distinct temperatures ($T_\parallel \ne T_\perp$),
and distinct power-indices ($\kappa\pa+s\pa \ne \kappa\se+s\se$).
We have introduced also $s\pa$
and $s\se$
because in the literature \cite{Lazar-etal-2012} different values for
$s\pa$
and $s\se$
are used. From the Table~\ref{tab:02} (see below) a choice of
$s\pa=s\se+1/2$
is appropriate to obtain symmetric parallel and perpendicular pressure
components. For a detailed discussion see below.  In the limit of
large parameters $\kappa, \kappa\pa, \kappa\se \rightarrow\infty$
the regularized $\kappa$ distributions approach the
corresponding Maxwellian (M) and Bi-Maxwellian (BM) distribution functions (see Table~\ref{tab:01})\\
\begin{table}[t!]
  \scalebox{0.85}{
\begin{tabular}{l|l}
  $f_{\bullet}$ & $\lim\limits_{\kappa, \kappa\pa, \kappa\se \rightarrow\infty}f_{\bullet}$\\
  \hline
  $f_{M}$ & $n_0 N_{M} \mathrm{e}^{-w^{2}}$\\
  $f_{K}$ & $n_0 N_{M} \mathrm{e}^{-w^{2}}=f_{M}$\\
  $f_{R}$ & $n_0 N_{R} e^{-(1+\alpha^{2}) w^2} \simeq n_0 N_{M} e^{-w^2} = f_{M}$ \\
  $f_{BK}$ & $n_{0}N_{BM} \mathrm{e}^{-w\pa^{2}-w\se^{2}}=f_{BM}$ \\
  $f_{\mathrm{PBK}}$ & $ n_{0}N_{BM}\mathrm{e}^{-w\pa^{2}-w\se^{2}}=f_{BM}$\\
  $f\su$ & $n_{0}N\su \mathrm{e}^{-(1+\alpha\pa^{2})w\pa^{2}-(1+\alpha\se^{2})w\se^{2}} 
	\simeq  n_{0}N_{BM}\ \mathrm{e}^{-w\pa^{2}-w\se^{2}} = f_{BM}$\\
  $f\pr$ & $n_{0}N\pr \mathrm{e}^{-(1+\alpha\pa^{2}) w\pa^{2}-(1+\alpha\se^{2}) w\se^{2}} 
	\simeq  n_{0}N_{BM}\ \mathrm{e}^{-w\pa^{2}-w\se^{2}} = f_{BM}$
\end{tabular}
}
\caption{\label{tab:01} The limits as $\kappa,\kappa\pa,\kappa\se\rightarrow\infty$ for
the regularized distribution functions.}
\end{table}
If $0<\alpha\pa, \alpha\se < 1$
are small enough, Maxwellian limits of $f\pr$
and $f\su$
reduce both to a standard bi-Maxwellian, i.e., $f_{BM}$,
with $N_{BM} = 1/(\pi^{3/2} \Theta_\parallel \Theta_\perp^2)$.
For isotropic temperature ($\Theta_\parallel = \Theta_\perp = \Theta$)
$f_{BM}$
reduces to an isotropic Maxwellian, $f_{M}$,
with $N_M= 1/ (\pi^{1/2} \Theta)^3$.
In the following we do not discuss the bi-$\kappa$
distribution further, because it behaves similar to the standard
$\kappa$-distribution, concerning the poles and higher order moments.

For drifting distributions we define a drift velocity
\begin{align}
    \vv{U}_{a}    &\equiv \Theta_{a} \vv{W}_{a} 
\end{align}
and all moments will depend then on the dimensionless quantities
$\vv{W},W\pa,\vv{W}\se$
with $\vv{W}=W\pa\vv{e}\pa+\vv{W}\se$,
with the appropriate normalization $\Theta,\Theta\pa,\Theta\se$
respectively. We neglect the argument of the moment $M$ (similar for the
tensor moments), when it is zero, i.e.\
$M(\vv{W}=\vv{0})\equiv M(\vv{0})\equiv M$.
This representation is particularly important in magnetized plasmas,
where the magnetic field imposes a preferential direction (parallel to
the magnetic field, subscript $\parallel$),
leading to gyrotropic distributions, which are isotropic in the plane
perpendicular (subscript $\perp$) to the magnetic field.

\section{Moments and most probable parameters}\label{sec:3}

The moments of the distribution functions are integrals over the
(entire) velocity space, where the volume elements are chosen
accordingly to avoid complicated integrations.  Thus, for the general
moments we use Cartesian normalized volume elements $\mathrm{d}^{3}v$,
while for the normalized isotropic distribution functions $f$
spherical volume elements $\vols$
are adopted, and for the anisotropic distribution functions we assume
cylindrical symmetry, i.e., gyrotropic distributions typical in
magnetized plasmas, with the corresponding normalized elements
$\volc$.

\subsection{General expressions for arbitrary non-drifting distribution functions}

We use the following general expressions for the principal moments (of
the zero, first, second and third orders) and the most probable values
defined for an arbitrary non-drifting distribution function
$f_{a}$ 
\begin{subequations}\label{genmom}
  \begin{align}\label{dens}
    n(\vv{r},t)\equiv M^{(0)}&= \int f(\vv{v},t)
    d^{3}v &\quad\mathrm{number\ density}\\\label{velo}
    \vv{u}(\vv{r},t)\equiv\vv{u}\equiv\vv{M}^{(1)}&= \dfrac{1}{n(\vv{r},t)}\int\vv{v} f(\vv{v},t)
    d^{3}v &\quad \mathrm{drift\ velocity}\\\label{speed}
    u_{p}(\vv{r},t)\equiv u_{p}\equiv M^{(1)}&= \dfrac{1}{n(\vv{r},t)}\int v f(\vv{v},t)
    d^{3}v &\quad \mathrm{most\ probable\ speed}\\\label{pres}
  \tensor{P}(\vv{r},t)\equiv\tensor{P}\equiv \tensor{M}^{(2)}&= \int\vv{v}\otimes\vv{v} f(\vv{v},t)
    d^{3}v &\quad \mathrm{pressure\ tensor}\\\label{heattens}
    \tensor{Q}(\vv{r},t)\equiv\tensor{Q}\equiv\tensor{M}^{(3)}&= \int\vv{v}\otimes \vv{v} \otimes\vv{v} f(\vv{v},t)
    d^{3}v &\quad \mathrm{heat \ flux\ tensor}\\\label{heatflux}
    \vv{q}(\vv{r},t)\equiv\vv{q}\equiv\vv{M}^{(3)}&= \int v^{2}\vv{v} f(\vv{v},t)
    d^{3}v &\quad \mathrm{heat \ flux\ vector}\\\label{heatave}
    q_{p}(\vv{r},t)\equiv q_{p}\equiv M^{(3)}&= \int v^{3} f(\vv{v},t)
    d^{3}v &\quad \mathrm{ most\ probable\ heat\ flux}
  \end{align}
\end{subequations}
$\otimes$
is the dyadic product of vectors or tensors and $v = |\vv{v}|$.  If the distribution
function $f$
is an even function in $w,w\pa,w\se$
(respectively in $v,v\pa,v\se$),
the integrals involving the products with odd functions of $\vv{w}$
(or equivalently $\vv{v}$)
vanish. Thus, without calculation one has $\vv{u}=\vv{0}$
and $\vv{q}=\vv{0}$
and $\tensor{Q}=\tensor{0}$.
Furthermore, in the pressure tensor only the elements in $\tr{P}\ne0$,
i.e.\ $P_{11},P_{22},P_{33}$
remain. Note: to get the correct physical units for the mass flow,
pressure, and heat flow the moments must be multiplied with the
particle mass.

The situation is different, when we allow for a velocity shift (i.e.\
a drift or bulk velocity) $\vv{U}=\Theta\vv{W}\ne0$
in the distribution function, i.e., $f(\vv{v}-\vv{U})$.
In order to evaluate the moments, in the distribution function we
replace $\vv{v}-\vv{U} \rightarrow \vv{v}^{\prime}$,
and, accordingly $v^{\prime2}=(\vv{v}-\vv{U})^{2}$,
and the integration variable to $\vv{v}=\vv{v}^{\prime}+\vv{U}$
and $v^{2}=(\vv{v}^{\prime}+\vv{U})^{2}$.
For the sake of simplicity we can drop ``prime'' and then find that
$n(\vv{r},t)$
remains unchanged, while the other moments become (with
$\Theta>0$)

\begin{subequations}\label{momW}
  \begin{align}\label{veloW}
    \vv{u}(\vv{W}) &= \vv{U}\\\label{speedW}
    u_{p}(\vv{W}) &= \dfrac{1}{n(\vv{r},t)}\int |\vv{v}+\vv{U}| f(\vv{v},t)
    d^{3}v \\\label{presW}
    \tensor{M}^{(2)}(\vv{W})&= \int(\vv{v}+\vv{U})\otimes(\vv{v}+\vv{U}) f(\vv{v},t)
                                       d^{3}v\\\nonumber
    &= \int(\vv{v}\otimes\vv{v}+\vv{U}\otimes\vv{U}) f(\vv{v},t)
    d^{3}v \\\label{heattensW}
    \tensor{M}^{(3)}(\vv{W})&= \int(\vv{v}+\vv{U})\otimes(\vv{v}+\vv{U})
                \otimes (\vv{v}+\vv{U})f(\vv{v},t) d^{3}v\\\nonumber
                &= \int(
                \vv{v}\otimes\vv{U}\otimes\vv{v}+
                \vv{v}\otimes\vv{v}\otimes\vv{U}+
                \vv{U}\otimes\vv{v}\otimes\vv{v}+
                \vv{U}\otimes\vv{U}\otimes\vv{U}
                )f(\vv{v},t)
    d^{3}v \\\label{heatflowW}
    \vv{q}(\vv{W})&= \int (\vv{v}+\vv{U})^{2}(\vv{v}+\vv{U}) f(\vv{v},t)
                              d^{3}v \\\nonumber
    &= \int (2(\vv{v}\cdot\vv{U})\vv{v}+v^{2}\vv{U}+U^{2}\vv{U}) f(\vv{v},t)
    d^{3}v\\\label{heataveW}
    q_{p}(\vv{W})&= \int |\vv{v}+\vv{U}|^{3} f(\vv{v},t)
    d^{3}v 
  \end{align}
\end{subequations}
where we have neglected all terms with an even times odd function.
The second order moment can be written
\begin{align}
    \tensor{M}^{(2)} = \tensor{P}+ n\tensor{UU}_{s}+n\tensor{UU}_{a}
\end{align}
where $\tensor{UU}_{s}$
is a symmetric tensor, with trace $\tr{UU}_{s}=U_{11}+U_{22}+U_{33}$
and all other elements vanish, while $\tensor{UU}_{a}$
is an antisymmetric tensor with $\tr{UU}_{s}=0$,
which describes the friction of the bulk speed. In a free flow we may
neglect the latter. We may identify $n\ \tr{UU}_{s}$
as the ``directional'' ram pressure of the bulk flow.  Moreover, the
zeroth, first, second and third order moment flow apparently have an analytic
solution, while the third moment, most probable speed and heat flux do
not have it in general. However, if we define the most probable
quantity along a parallel or perpendicular direction, an analytic
solution can be found and is given below.

\subsection{Non-drifting distributions: $\vv{W}=\vv{0}$}
\begin{table}
  \begin{tabular}{l|l|l|l|}
    &$f_{R}$
&$f_{K}$
&$f_{M}$
\\
& $(\kappa,\alpha)$
& $(\kappa)$
& $$
\\
  \hline
  $N$
  & $\frac{1}{\sqrt{\pi^{3}\kappa^{3} }\Theta^{3}}\ \Ika[][0]$
  & $\frac{\GA{\kappa}}{\Theta^{3}\sqrt{\pi^{3}}\sqrt{\kappa^{3}}\GA{\kappa-\frac{1}{2}}}$
  & $\frac{1}{\Theta^{3}\sqrt{\pi^{3}}}$
  \\
  \hline
  $\vv{u}$
  & $\vv{0}$
  & $\vv{0}$
  & $\vv{0}$
  \\
  \hline
  $u_{p}$
  &$\frac{2}{\sqrt{\pi}}\Theta\sqrt{\kappa}\Ika[1][0]$
  &$\frac{2}{\sqrt{\pi}}\Theta\sqrt{\kappa}\frac{\GA{\kappa-1}}{\GA{\kappa-\frac{1}{2}}}$
  & $\frac{2}{\sqrt{\pi}}\Theta$
  \\
  \hline
  $P_{11}$
  &$\frac{1}{2}n_{0}\Theta^{2}\kappa \Ika[2][0]$
  &$n_{0}\Theta^{2}\frac{\kappa}{2\kappa-3}$
  &$\frac{1}{2} n_{0} \Theta^{2}$
  \\
  \hline
  $P_{22}=P_{33}$
  &$P_{11,R}$
  &$P_{11,K}$
  &$P_{11,M}$
  \\
  \hline
  $p_{ij},i\ne j$
  & $0$
  & $0$
  & $0$
  \\
  \hline
  $\vv{q}$
  & $\vv{0}$
  & $\vv{0}$
  & $\vv{0}$
  \\
  \hline
  $q_{p}$
  &$\frac{4}{\sqrt{\pi}}n_{0}\Theta^{3}\sqrt{\kappa^{3}} \Ika[3][0]$
  &$\frac{4}{\sqrt{\pi}}n_{0}\Theta^{3}\sqrt{\kappa^{3}}\frac{\GA{\kappa-2}}{\GA{\kappa-\frac{1}{2}}}$
  &$\frac{4}{\sqrt{\pi}}n_{0}\Theta^{3}$
  \\
  \hline
  $\tensor{Q}$
  & $\tensor{0}$
  & $\tensor{0}$
  & $\tensor{0}$
  \\

   \end{tabular}
   \caption{ \label{tab:02a} The moments for non-drifting isotropic
     distribution functions including the standard
     $\kappa$-distribution.
     }
\end{table}

The details of the calculation of the moments are given in the appendices,
as explained below. The results are found in Table.~\ref{tab:02a},
where we have introduced the functions $\Ika[n][m](\kappa,\alpha)$;
\begin{align}
  \Ika[m][n](\kappa,\alpha) =
  \frac{U\left(\frac{3+m}{2},\frac{3+m}{2}-\kappa, \alpha^{2}\kappa\right)}
  {U\left(\frac{3+n}{2},\frac{3+n}{2}-\kappa, \alpha^{2}\kappa\right)}
\end{align}
The indices $n$
denote the numerator of the velocity integral, where the integrand is
proportional to $v^{n}$,
while $m$
is analogous for the denominator. The denominator is in principal the
normalisation, while the numerator describes the order of the moment
(except for the RPBK distribution, see Appendix~\ref{appInt} for
details). The function \Ika[m][n] can be expressed as the ratio of two
Kummer~U (or two Tricomi) functions.  Thus, for example, the pressure
of the regularized $\kappa$-distribution
is proportional to \Ika[2][0] (see Table~\ref{tab:02a}).

In table ~\ref{tab:02} the moments for all non-drifting ani\-sotropic
distribution discussed above are given (for an explicit calculations
see Appendix~\ref{appInt}).  The function
$\Jka[n,m][l,k](\kappa,\ap,\as)$
is defined in a similar way like the function $\Ika[][]$,
but the last integral has to be solved numerically (see
Eqs.~\ref{numer},~\ref{mrpbk} and~\ref{mrpbke}).

The indices $n,m$
and $k,l$
are for the parallel and perpendicular direction respectively. For the
PBK we had to introduce two slightly different function \Vka[m][n][\pa]
and \Vka[m][n][\se] (Eqs.~\ref{eq:tripro} and~\ref{eq:prodmom}).  From
these equations the moments of the other distributions (K, BK, and
PBK) can easily be  derived using Eqs.~\ref{eq:K},~\ref{eq:BK}
and~\ref{eq:PBK}
\subsection{Drifting distributions: $\vv{W}\ne\vv{0}$}

The velocity and pressure moments are discussed above and are
generally given by
\begin{eqnarray}
  \vv{u}= \Theta\vv{W} \qquad \tensor{M}^{(2)} = \tensor{P} + n \tensor{UU}.
\end{eqnarray}
Here we calculate the heat flow vector, and the most probable speeds
and heat flows. The heat flow vectors for distribution functions with
isotropic temperatures ($A=1$) take the following form (see
Eq.~\ref{eq:heatdrift})
\begin{align}\label{eq:drk}
  \vv{q} &=
  \left(P_{11} + n ?Theta^{2}W^{2}\right)\Theta\vec{W}.
\end{align}
For the anisotropic temperatures we obtain (see Eq.~\ref{heataniso})
\begin{align}\label{eq:dbk}
  \vec{q} = & 2
  ( P_{11}  \Tp W\pa,
   P_{22}  \Ts W_{\se,1},
   P_{22}   \Ts W_{\se,2})^{T}
  +[ P_{11}+ P_{22}]\ 
    (\Theta\pa W\pa, \Theta\se W_{\se,1} ,\Theta\se W_{\se,2})^{T}
   +\\\nonumber
  &\hspace*{6cm} n  [\Theta\pa^{2} W^{2}\pa +
  \Theta^{2}\se W\se^{2}]\ 
 (   \Tp W\pa , \Ts W_{\se,1} , \Ts W_{\se,2})^{T}
\end{align}

The integrals are more complicated, when we treat the modulus of the
velocity and the heat flow, i.e., the most probable speeds and the
most probable heat flows, which are discussed in the following.

\clearpage

\begin{sidewaystable}
  \scalebox{1.1}{
    \begin{tabular}{l|l|l|l|l|l|l}
      \multicolumn{7}{c}{~}\\
      \multicolumn{7}{c}{~}\\
      \multicolumn{7}{c}{~}\\
      \multicolumn{7}{c}{~}\\
      \multicolumn{7}{c}{~}\\
      \multicolumn{7}{c}{~}\\
      \multicolumn{7}{c}{~}\\
      \multicolumn{7}{c}{~}\\
      \multicolumn{7}{c}{~}\\
      &$f_{\mathrm{PBK}}$
&$f\pr$
&$f_{\mathrm{BK}}$
&$f_{\mathrm{RBK}}$
&$f_{\mathrm{RBK}}$
&$f_{BM}$
\\
& $(\kappa\pa,s\pa),(\kappa\se,s\se)$
& $(\kappa,\alpha\pa,s\pa),(\kappa,\alpha\se,s\se)$
& $(\kappa)$
& $(\kappa,\alpha\pa,\alpha\se)$
&$(\kappa,\alpha\pa=\alpha\se)$
&$$
\\
  \hline
  $N$
  &
  $\frac{\GA{\kappa\pa+s\pa}(\kappa\se+s\se-1)}{\sqrt{\pi^{3}}\Theta\pa\Theta\se^{2}
    \sqrt{\kappa\pa}\kappa\se
    \GA{\kappa\pa+s\pa-\frac{1}{2}}}$
  &
  $\frac{\Vka[][0][s\pa]\Vka[][0][s\se]}{\sqrt{\pi^{3}}\Theta\pa
    \Theta^{2}\se\sqrt{\kappa_{\pa}}\kappa_{\se}}            $
  & $\frac{\GA{\kappa}}{\sqrt{\pi^{3}\kappa}\Theta_{\pa}\Theta_{\se}^{2}\GA{\kappa-\frac{1}{2}}}$
  & $\frac{1}{\sqrt{\pi^{3}\kappa^{3}}\Theta\pa \Theta^{2}\se} \Jka[][0,0]$
  & $\frac{1}{\sqrt{\pi^{3}\kappa^{3}}\Theta\pa\Theta\se^{2}}\Ika[0]$
  & $\frac{1}{\sqrt{\pi^{3}}\Theta\pa\Theta\se^{2} }$  
  \\
  \hline
  $\vv{u}$
  & $\vv{0}$
  & $\vv{0}$
  & $\vv{0}$
  & $\vv{0}$
  & $\vv{0}$
  & $\vv{0}$
  \\
  \hline
  $u_{p\pa}$
  &$\frac{1}{\sqrt{\pi}}\Theta\pa \frac{\sqrt{\kappa_{\pa}}\GA{\kappa\pa+s\pa-1}}{\GA{\kappa\pa+s\pa-\frac{1}{2}}}$
  &$\frac{1}{\sqrt{\pi}}\Theta\pa \sqrt{\kappa_{\pa}}\Vka[1][0][\pa]$
  &$2\Theta_{\pa}\sqrt{\kappa}\frac{\GA{\kappa-1}}{\GA{\kappa-\frac{1}{2}}}$
  & $\frac{2}{\sqrt{\pi}}\Theta\pa\sqrt{\kappa}\Jka[1,0][0,0]$
  & $\frac{2}{\sqrt{\pi}} \Theta\pa\sqrt{\kappa}\Ika[1][0]$
  & $\frac{1}{\sqrt{\pi}}\Theta\pa$
  \\
  \hline
  $u_{p\se}$
  &$\frac{\sqrt{\pi}}{2} \Theta\se\frac{\GA{\kappa\se+s\se-\frac{3}{2}}}{\GA{\kappa\se+s\se-1}}$
  &$\frac{1}{2\sqrt{\pi}}\Theta\se\sqrt{\kappa\se} \Vka[1][0][\se]$
  &$\frac{2}{\sqrt{\pi}}\Theta_{\se}\sqrt{\kappa}\frac{\GA{\kappa-1}}{\GA{\kappa-\frac{1}{2}}}$
  &$\frac{2}{\sqrt{\pi}} \Theta\se\sqrt{\kappa}\Jka[0,1][0,0]$
  &$\frac{2}{\sqrt{\pi}} \Theta\se\sqrt{\kappa}\Ika[1][0]$
  &$\frac{1}{2}\sqrt{\pi}\Theta\se$
  \\
  \hline
  $P_{11}$
  &$n\Theta^{2}\pa \frac{\kappa\pa}{2(\kappa_{\pa}+s\pa)-3}$
  &$\frac{1}{2} n \Theta^{2}\pa \kappa\pa \Vka[2][0][\pa]$
  &$n\Theta_{\pa}^{2}\,\frac{\kappa}{2\kappa-3}$
  &$\frac{3}{2} n \Theta^{2}\pa \kappa \Jka[2,0][0,0]$
  &$\frac{1}{2}      n \Theta^{2}\pa \kappa \Ika[2][0]$
  &$ \frac{1}{2} n \Theta^{2}\pa$
  \\
  \hline
  $P_{22}=P_{33}$
  &$ n\Theta^{2}_{\se}\frac{\kappa\se}{2(\kappa_{\se}+s\se)-4}$
  &$ \frac{1}{2} n \Theta^{2}_{\se} \kappa\se \Vka[2][0][\se]$
  &$n \Theta^{2}_{\se}\,\frac{\kappa}{2\kappa-3}$
  &$\frac{3}{4} n \Theta^{2}_{\se}\kappa \Jka[0,2][0,0]$
  &$  \frac{1}{2}         n \Theta^{2}_{\se}\kappa \Ika[2][0]$
  &$\frac{1}{2} n \Theta\se^{2}$
  \\
  \hline
  $p_{ij},i\ne j$
  & $0$
  & $0$
  & $0$
  & $0$
  & $0$
  & $0$
  \\
  \hline
  $\vv{q}$
  & $\vv{0}$
  & $\vv{0}$
  & $\vv{0}$
  & $\vv{0}$
  & $\vv{0}$
  & $\vv{0}$
  \\
  \hline
  $q_{p\pa}$
  &$\frac{1}{\sqrt{\pi}}n\Theta^{3}\pa\sqrt{\kappa\pa^{3}}\frac{\GA{\kappa\pa+s\pa-2}}{\GA{\kappa\pa+s\pa-\frac{1}{2}}} $
  & $\frac{1}{\sqrt{\pi}}n\Theta^{3}\pa\sqrt{\kappa_{\pa}^{3}} \Vka[3][0][s\pa]$
  &$\frac{2}{\sqrt{\pi}}n \Theta\pa^{3}\sqrt{\kappa^{3}}\frac{\GA{\kappa-2}}{\GA{\kappa-\frac{1}{2}}}$
  &$\frac{4}{\sqrt{\pi}} n \Theta\pa^{3}\sqrt{\kappa^{3}} \Jka[3,0][0,0]$
  &$\frac{2}{\sqrt{\pi}}n \Theta\pa^{3}\sqrt{\kappa^{3}} \Ika[3][0]$
  &$\frac{1}{\sqrt{\pi}}n \Theta\pa^{3}$
  \\
  \hline
  $q_{p\se}$
  &$ \frac{3\sqrt{\pi}}{4}n\Theta^{3}\se\sqrt{\kappa^{3}}\frac{\GA{\kappa\se+s\se-\frac{5}{2}}}{\GA{\kappa\se+s\se-1}}$ 
  &$\frac{3\sqrt{\pi}}{4} n \Theta^{3}\se\sqrt{\kappa^{3}} \Vka[3][0][s\se]$ 
  &$\frac{3\sqrt{\pi}}{2}n \Theta_{\se}^{3}\sqrt{\kappa^{3}}\frac{\GA{\kappa-2}}{2\GA{\kappa-\frac{1}{2}}}$
  &$\frac{3\sqrt{\pi}}{2}n\Theta^{3}\se\sqrt{\kappa^{3}}
  \Jka[0,3][0,0]$
  &$\frac{3\sqrt{\pi}}{2}n\Theta^{3}\se\sqrt{\kappa^{3}} \Ika[3][0]$
  &$\frac{3}{4}\sqrt{\pi}n\Theta^{3}\se$
  \\
  \hline
  $\tensor{Q}$
  & $\tensor{0}$
  & $\tensor{0}$
  & $\tensor{0}$
  & $\tensor{0}$
  & $\tensor{0}$
  & $\tensor{0}$
  \\

  \end{tabular}
  }
  \caption{ \label{tab:02} The moments  for
    non-drifting distribution functions. Column~7 provides the moments for the bi-Maxwellian. }
\end{sidewaystable}
\clearpage

\noindent

\subsubsection{The most probable parallel speed and heat flow}

The most probable parallel speed for anisotropic distribution
functions can be written as (see Appendix~\ref{appMost}, Eqs.~\ref{eq:mpswp} and~\ref{eq:mphwp}):
\begin{align}
  u_{p\pa}(W\pa) &=  u_{p\pa}(0) +\Theta\pa y_{i} (W\pa) \\
   q_{p\pa}(W\pa) &=  q_{p\pa} + 3 \Tp
    W\pa P_{22p\pa}(\vv{0}) +2 n \Theta^{3}\pa z_{i}
\end{align}
and for the RPBK and RBK distributions $y_{i}$
and $z_{i}$ are given in Table~\ref{tab:03} and Table~\ref{tab:04}.

\begin{table*}[t]
\begin{tabular}{l|l}
  & $y_{i}(W\pa)$\\
\hline
  $u_{p,\mathrm{BM}\pa}$ &
                 $\frac{W_{\pa}}{\ap} \erf(W_{\pa})$ \\
  $u_{p,\mathrm{RPBK}\pa}$ & $\frac{1}{2}W\pa \Vka[][0][\pa](\kappa\pa,\alpha\pa)\left\{\ika[1][](\kappa\pa,\alpha\pa) +\ika[0][](\kappa\pa,\alpha\pa)\right\}$ \\
  $u_{p,\mathrm{RBK}\pa}$ &
                      $\frac{1}{2}W\pa \Jka[][0,0](\kappa\pa,\kappa\se,\alpha\pa,\alpha\se)
                      \left\{\jka[1,0][]^{\pa}(\kappa\pa,\kappa\se,\alpha\pa,\alpha\se)+
                       \jka[1,0][]^{\pa}(\kappa\pa,\kappa\se,\alpha\pa,\alpha\se)\right\}$ 
\end{tabular}
\caption{\label{tab:03}The parameter $y_{i}$ for the most probable parallel speed.}
\end{table*}

\begin{table*}[t]
\begin{tabular}{l|l}
  & $z_{i}(W\pa)$\\
  \hline
    $q_{p,\mathrm{BM}\pa}$ & $
                   \frac{1}{2\sqrt{\pi}}\left(\frac{1}{\pi\ap^{3}}+\frac{3W\pa^{2}}{\ap}\right)
                   \left(\mathrm{e}^{-\ap^{2}W\pa^{2}}-1\right) +
                   \frac{W\pa\erf(\ap W\pa)}{2\pi\ap^{2}}\left(\frac{3}{2}+2\ap^{2}W\pa^{2}\right)$\\
    $q_{p,\mathrm{RPBK}\pa}$ & $\Vka[][1][\pa](\kappa\pa,\ap)\left(W\pa^{3}\ika[0][](\kappa\pa,\ap)
                     -3 W\pa^{2} \ika[1][](\kappa\pa,\ap)+3
                     W\pa \ika[2][](\kappa\pa,\ap)-\ika[3][](\kappa\pa,\ap)\right)$\\
    $q_{p,\mathrm{RBK}\pa}$ & $\Jka[][0,0](\kappa\pa,\kappa\se,\ap,\as) \left(W\pa^{3}\jka[0,0]^{\pa}(\kappa\pa,\kappa\se,\ap,\as)
                     -3 W\pa^{2}
                        \jka[1,0]^{\pa}(\kappa\pa,\kappa\se,\ap,\as)
                        \right.$\\
  &\hspace{4cm}$\left.+3
                     W\pa \jka[2,0]^{\pa}(\kappa\pa,\kappa\se,\ap,\as)-\jka[3,0]^{\pa}(\kappa\pa,\kappa\se,\ap,\as)\right) $
\end{tabular}\\
\caption{\label{tab:04}The $z_{i}$ parameter for the most probable parallel heat flow.}
\end{table*}

~\\~\\
\paragraph{The case $|W\pa| \ll 1$}
The values $y$
and $z$
can be assumed negligible for small values of $|W\pa|\ll 1$
and $0<\ap \ll1$,
implying that we can simplify \ika[n][], then the required integrals
have values between $\{0\ldots |W\pa|\}$
and $|w\pa|$ is also small. Thus, we may approximate the exponential
in Eq.~\ref{eq:mpswp}
\begin{align}
  &\left(1+\frac{w^{2}\pa}{\kappa}\right)^{-\kappa-1}
  \mathrm{e}^{-\ap^{2} w^{2}} \approx
\left(1+\frac{w^{2}\pa}{\kappa}\right)^{-\kappa-1}\left(
  1-\ap^{2}w^{2}\pa+\frac{\ap^{4}w^{4}\pa}{2} \ldots\right)
\end{align}
and the leading term of the functions \ika[i][] can easily be
calculated as
\begin{align}\nonumber
 \hspace*{-0.5cm} \ika[0][](\kappa\pa,\alpha\pa) &\approx W\pa \hyp[2][1]\left(\left[\frac{1}{2},\kappa+1\right],\left[\frac{3}{2}\right],-\frac{W\pa^{2}}{\kappa}\right)\ldots\\\nonumber
 \hspace*{-0.5cm} \ika[1][](\kappa\pa,\alpha\pa) &\approx \frac{1}{2}-\frac{1}{2}\kappa^{\kappa}(W\pa^{2}+\kappa)^{-\kappa}\ldots\\\nonumber
 \hspace*{-0.5cm} \ika[2][](\kappa\pa,\alpha\pa) &\approx W\pa^{3} \hyp[2][1]\left(\left[\frac{3}{2},\kappa+1\right],\left[\frac{5}{2}\right],-\frac{W\pa^{2}}{\kappa}\right)\ldots\\\nonumber
 \hspace*{-0.5cm} \ika[3][](\kappa\pa,\alpha\pa) &\approx \frac{\kappa}{2(\kappa-1)}
                                   \left\{1 - \kappa^{\kappa}(W\pa^{2}+1)(W\pa^{2}+\kappa)^{-\kappa} \right\}\ldots\
\end{align}

A similar approximation for the \jka[i,m][], leads to (see
Appendix~\ref{appInt}, Eq.\ref{numer})
\begin{align}
  \jka[i,1]^{\parallel} \approx
  \frac{1}{2}\mathrm{e}^{\as^{2}\kappa}\kappa^{\kappa}\as^{2\kappa}\Gamma(-\kappa,\as^{2}\kappa)
  \ap^{i+1} \hyp[2][1]\left(\left[\frac{i+1}{2},\kappa+1\right],\left[\frac{i+3}{2}\right],-\frac{W\pa^{2}}{\kappa}\right)
\end{align}
for $i\in\{0,1,2,3\}$. The Taylor expansion for $z_{\mathrm{RPBK}}$ gives
\begin{align}
  z_{\mathrm{RPBK}} \approx \frac{W\pa^{4}}{4}
\end{align}
Thus, the values $y_{\mathrm{RPBK}},z_{\mathrm{RPBK}}$
are negligible if the parallel drift speed $W\pa$ is small.

\subsubsection{The perpendicular most probable speed}

The most probable speeds and heat flow are calculated in the
Appendix~\ref{appMost}.  For a constant drift velocity $\vv{W}$
the most probable speed and heat flow are given by:
\begin{align}
  u_{p\se}(W\se) &=  u_{p\se} (\vv{0})+  n \Tp \Ts^{3}
                       W\se^{2} N \mathcal{I}_{0} \\
  q_{p\se}(W\se) &=  q_{p\se} (\vv{0}) + 
                       2\pi\Theta^{2}\se[1  + W_{\se}^{2}]
    u_{p\se}(\vv{0})+n \Tp \Ts^{5}W\se^{4}N  \mathcal{I}_{0} 
\end{align}
where $\mathcal{I}_{0}$ is given in Table~\ref{tab:05}.
\begin{table}
\begin{tabular}{l|l}
  & $\mathcal{I}_{0}$\\
  \hline
  $f_{\mathrm{BM}\se}$      & $ \frac{\sqrt{\pi^{3}}}{2\ap\as}$ \\
  $f_{\mathrm{RPBK}\se}$    & $\Vka[0][1][\se](\kappa\se,\as) $\\
  $f_{\mathrm{RBK}\se}$ &$\Jka[0,0][0,1](\kappa\pa.\kappa\se,\ap,\as) $
\end{tabular}
\caption{\label{tab:05}The most probable perpendicular speed.}
\end{table}
In the special case $\vv{W}_{w\se} = a \frac{\vv{w}\se}{w\se}$
with $a\in\mathbb{R}$,
we can use Eq.~\ref{s1} and~\ref{q1} to evaluate the most probable
speed and heat flow to
\begin{align}\nonumber
   u_{p\se}(\vv{W}_{w\se}) &= u_{p\se}(\vv{0}) + a \Theta\se
                                  + a^{2} \Tp\Ts^{3}N_{} \mathcal{I}_{0}\\\nonumber
    q_{p\se}(\vv{W}_{w\se}) &= q_{p\se}(\vv{0}) + a\Theta\se
                                  P_{22\se}(\vv{0}) +a^{2}\Theta\se^{2}u_{p\se}(\vv{0})
    + a^{3}n \Theta\se^{3} +  a^{4}n \Tp\Ts^{5} N \mathcal{I}_{0}
\end{align}
The above assumption about $\vv{W}_{w\se}$
can be applied for shrinking or expanding perpendicular dynamics. But
a thorough discussion would go far beyond the scope of this paper.  In
the general case, where
$\vec{w}_{\se} = \vec{w}_{\se 1}+\vec{w}_{\se 2}$
no analytic solution was found (see appendix~\ref{appMost}).

\section{Illustration and discussion \label{sec:4}}
 We have calculated in Table~\ref{tab:01} and~\ref{tab:02a} the
  pressure components $P_{11},P_{22},P_{33}$.
  Now we want to describe the components $P\pa$
  and $P\se$,
  i.e., the parallel and perpendicular pressure components as well as
  the respective temperature components. For the parallel temperature
  we use the classical definition, with temperature defined as the
  average  kinetic energy
\begin{align}
  \kB T_{\pa} &=
                \frac{m}{n_{0}}\int\limits_{-\infty}^{\infty}v\pa^{2}f(v\pa)\mathrm{d}v\pa= \frac{m}{n_{0}} P_{11}\equiv \frac{m}{n_{0}} P\pa  
\end{align}
where we have used the decomposition
$\vec{v}=v\pa\vec{e}\pa+v_{\se,1}\vec{e}_{\se,1}+v_{\se,2}\vec{e}_{\se,2}$
and the factor $1/n_{0}$
to get the correct physical units.
 Here we assume that the perpendicular vectors have the same
  components in both directions with the unit vectors
  $\vec{e}_{\se,1,2}$.
  In general $\vec{v}_{\se}$ does not
  need to have the same value in direction $\vec{e}_{\se,1}$
  and $\vec{e}_{\se,2}$
  which leads to a fully anisotropic distribution function
  $v_{\pa}\ne v_{\se,1}\ne v_{\se,2}$,
  especially the perpendicular temperatures and pressures are
  also anisotropic
  (see Effenberger et al. 2012b, for a discussion of cosmic ray
  diffusion). Usually, one is only interested
  in a decomposition of, say, the magnetic field in a parallel and
  isotropic perpendicular components. A more general decomposition of
  the velocity components would also lead to a more general (R)BK
  where $v\se$
  has to be decomposed, then a full 3D Cartesian integration with
  constants $\Theta_{x},\Theta_{y},\Theta_{z}$
  and $\alpha_{x},\alpha_{y},\alpha_{z}$
  has to be carried out. For most purposes  is sufficient to assume an
  isotropic perpendicular distribution functions and use only
  the amplitudes $v_{\pa}$ and $v_{\se}$.  

This leads for the bi-Maxwellian
distribution (BM) to the classical expression
$\kB T\pa=m\Theta\pa^{2}/2$,
where $\Theta\pa$
is the classical thermal speed (see Table~\ref{tab:02a}).  Now
defining the perpendicular temperature, one needs to be a little more
careful, because the integration is now over $\vec{v}\se^{2}$
\begin{align}
  \kB T_{\se} &\equiv
                \frac{m}{2n_{0}}\int\limits_{0}^{\infty}(\vec{v}_{\se,1}^{2}+\vec{v}_{\se,2}^{2})f(v\se)v\se\mathrm{d}v\se= \frac{m}{2n_{0}} (P_{22}+P_{33})=\frac{m}{n_{0}}P_{22}
    =\frac{m}{n_{0}}P_{33} 
  \equiv \frac{m}{n_{0}} P\se  
\end{align}
the latter identities hold with the assumption of a gyrotropic
distribution function (e.g.\ in magnetized plasmas)
$P_{22}=P_{33}$.

The above result holds true for all distribution functions
  discussed here, even if they are not separable in the integrals
  discussed below. For the pressure (temperature) it turns out that
  the parallel and perpendicular components are mutually
  independent. This may hold true for more complicated distribution functions.


\subsection{Temperature  anisotropy}

We do not discuss the details of the RPBK or PBK because in the limit
of $\kappa\pa=\kappa\se$
and $\Tp=\Ts$
they do not approach the isotropic case. This is due to the fact, that
when multiplying the two factors in Eq.~\ref{eq:rpbk} (for
$\kappa\pa+s\pa=\kappa\se+s\se$)
we are always left with a bi-quadratic term $v\pa^{2}v\se^{2}$.
In the literature often $s\pa=s\se=1$
is used \cite[e.g.][and references therein]{Lazar-etal-2012} which
leads to an asymmetric expression for the pressures $P_{11}$ and
$P_{22}$, while these terms for the other discussed distribution
functions (see Table~\ref{tab:02}) are symmetric. We can heal this
behavior by choosing $s\pa=s\se+1/2$ which leads to 
$\kappa\se+s\se=\kappa\se+s\pa+1/2$.

\begin{figure}[t!]
  \includegraphics[width=0.95\textwidth]{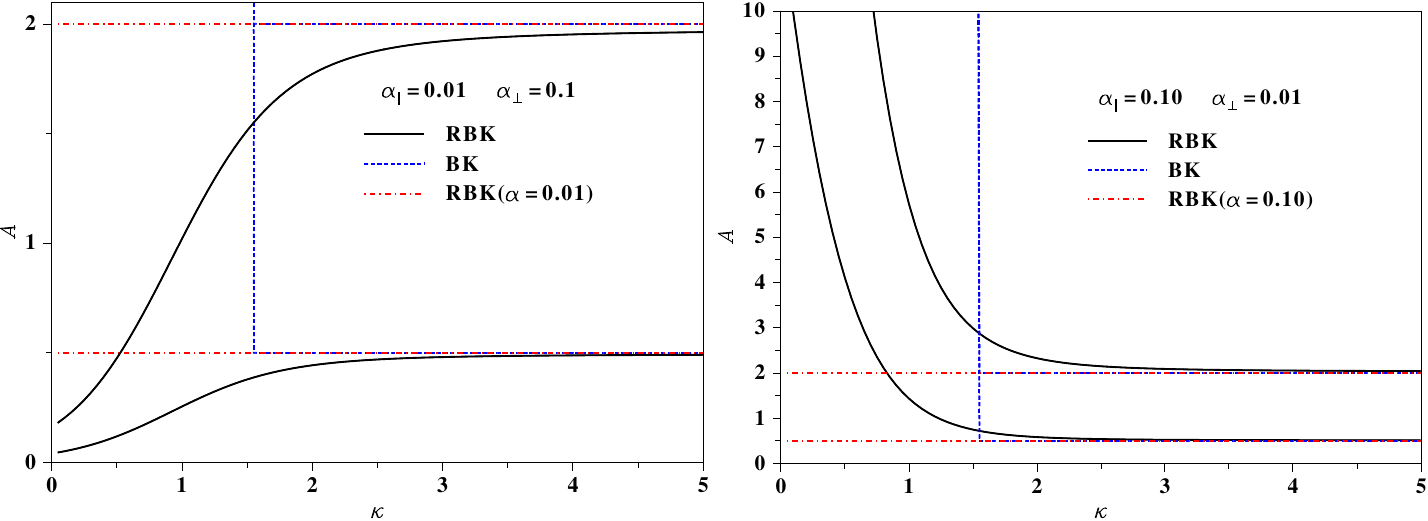}
  \caption{The anisotropy $A$. In the left panel the anisotropy
    for $\ap=0.01$ and $\as=0.1$ is shown and vice versa in the right
    panel. The anisotropy is chosen in such a way that for large
    $\kappa$ values $A=\Ts^{2}/\Tp^{2}$ has a constant value, for the
    upper curves this is $A=2$ and for the lower curves
    $A=1/2$. For more details see text.
    \label{fig:f1}}
\end{figure}

In Fig.~\ref{fig:f1} we have in both panels plotted the anisotropy
$A=T\se/T\pa$ which, in the limit $\kappa\to\infty$, becomes either $A=\Ts^{2}/\Tp^{2}= 2$ or $A=1/2$
 for our two cases discussed in Fig.~\ref{fig:f1}. 
The red curves denote the case when in the RKB $\alpha=\as=\ap$,
as one can see the anisotropy remains at the same value (for
$A = 2$
the red upper curves in both panels and for $A=1/2$
the lower red curves.) The BK is given by the blue curves, which are
identical to the red ones in the range $\kappa>1.5$,
below that values the BK is not defined and at $\kappa=1.5$
it is infinite. The interesting case are the black curves, which
change the anisotropy for small $\kappa<1.5$.
In the left panel the cutoff parameters are $\ap=0.01$
and $\as=0.1$
and in the right panel $\ap=0.1$
and $\as=0.01$.
The interesting case is that, as one can see in the left panel, the
anisotropy, which is $A=2$
for high $\kappa$
values, changes its behavior from a $A>1 $
to $A<1$
with an intersection point $\kappa_{s}$
where $A_{s}=1$
and the anisotropy vanishes. Thus, for $A=2$
the perpendicular temperature is higher than the parallel one, both
become equal at the critical point $\kappa_{s}$, and for smaller
$\kappa<\kappa_{s}$ the parallel temperature becomes higher than the
perpendicular one. 

In the right panel of Fig.~\ref{fig:f1} ($\ap=0.1, \as=0.01$)
the anisotropy increases for small $\kappa$
values to very large values. Especially, in the case $A=1/2$
it becomes higher than one and, thus, the perpendicular temperature
becomes higher than the parallel one.  This behavior can be explained,
because the stronger cutoff $\ap>\as$
(i.e.\ less high speed particles) contributes to the pressure and that
causes the anisotropy variation for small $\kappa$'s,
while for larger $\kappa$
values the distribution functions become more Maxwellian, and thus
suppress the high speed contributions. The above described feature
needs further discussions and is especially interesting for the
stability of anisotropic plasmas
\cite[e.g.][and references therein]{Shaaban-etal-2019}.

The RBK distribution for $\ap=\as$ (red curves) has a constant anisotropy which
does not change with $\kappa$. The behavior of the BK-distribution is
similar, except that anisotropy cannot be definded for  $\kappa\le 3/2$.

\subsection{The heat flow vector for the drifting RK and RBK distributions}

\begin{figure}[t!]
  \includegraphics[width=0.45\textwidth]{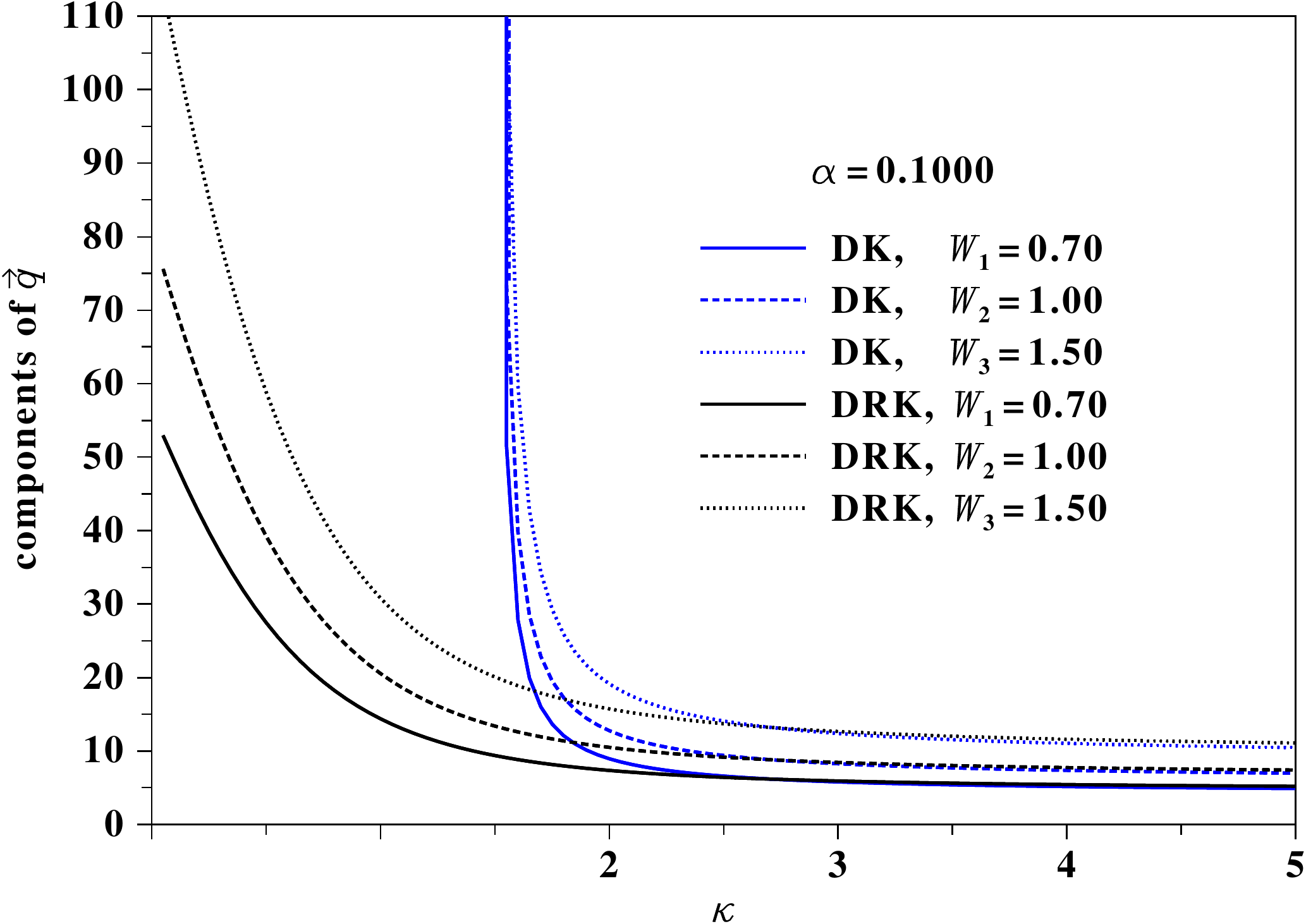}
  \caption{The  heat flow for the DK-
    (blue) and the DRK-distributions (black) is shown
    for the components of the drift vector
    $\vec{W}=(W_{1},W_{2},W_{3})^{T}$. The curves only differ by the
    values of the components of the drift vector $\vec{W}$. See text
    for a discussion.
    \label{fig:f2}}
\end{figure}

\begin{figure}[t!]
  \includegraphics[width=0.95\textwidth]{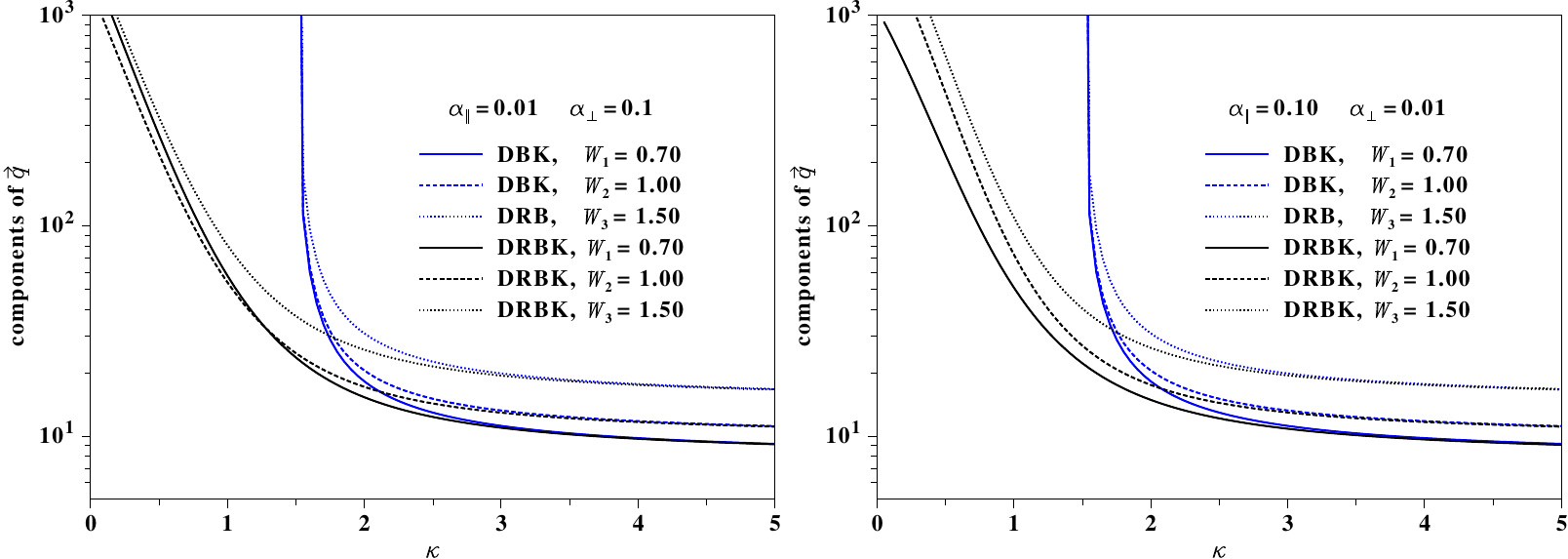}
  \caption{Similar as in Fig.~\ref{fig:f2} for the anisotropic heat
    flow for the DBK- (blue) and DRBK-distribution (black). The left
    panel is for the cutoff parameter $\ap=0.01$
    and $\as=0.1$ and the lower one for $\ap=0.01$ and $\as=0.1$
    \label{fig:f3}}
\end{figure}

In Fig.~\ref{fig:f2} and~\ref{fig:f3} the heat flow is calculated for
the drifting K and RK distributions (DK and DRK) as well as the BK and
RBK distributions (DBK and DRBK) for the same drift vector
$\vec{W}=(0.7,1,1.5)^{T}=(W\pa,W_{\se,1},W_{\se,2})^{T}$,
according to Eq.~\ref{eq:drk} for the isotropic and to
Eq.~\ref{eq:dbk} for the anisotropic temperatures.  In
Fig.~\ref{fig:f2} it can be seen that the heat flow is a constant
factor for all three components (see Eq.~\ref{eq:drk}) and, thus,
depends strongly on the values of the drift vector $\vec{W}$
for both the DK (blue) and DRK (black) distributions.

The heat flows obtained in the case of DRBK- and DBK-distributions are
shown in Fig.~\ref{fig:f3} as black and blue curves, respectively. In
both panels $A=1/2$.
They have a more interesting feature: As can be seen in the left panel
(for $\ap=0.01, \as=0.1$)
the parallel component $W(1)$
intersects one of the perpendicular components, i.e., $W(2)$
and marginally for very small $\kappa$-values
touches the $W(3)$
component. In the right panel (for $\ap=0.1, \as=0.01$)
the curves do not intersect, but are obviously not parallel for small
$\kappa$
values. If the heat flow components intersect for $\kappa<2$
depends on the choice of the drift vector components.

Thus, if  $W_{\se,1} \ne W_{\se,2}$ the drift in the two
perpendicular directions differs and and one can expect non-isotropic
turbulence or more complex diffusion tensors for cosmic ray
propagation \cite{Effenberger-etal-2012,Effenberger-etal-2012b}.  
Again, this behavior needs further research and comparison with
data. But this is not the goal of this work.

\section{Conclusions and perspectives \label{sec:5}}

We have introduced new regularized forms for the anisotropic
$\kappa$-distributions
which can reproduce temperature anisotropies (i.e., the regularized
bi-$\kappa$
(RBK) and the regularized product bi-$\kappa$
(PBK) distributions) and arbitrary drifts or flow speeds.  We have
shown that these distributions admit all higher order moments, which
are well defined for all values of $\kappa$.
In section~III we have estimated these moments e.g.\ the pressure and
heat flux tensors, as well as the heat flow vector. For an
illustration, in section~IV we have discussed the parallel and
perpendicular components of the temperature (pressure) and their
anisotropies. In addition we have also estimated the heat flow
components for representative drifting distributions, i.e.\ DRBK and
DRPBK by contrast with the DBK and DPBK.  The case of DRPBK is not
discussed in detail because of the problems to recover the standard
BK for isotropic temperatures.

For RBK and BK we discussed the general case, when the cutoff
parameters $\alpha_{\pa}\ne\alpha_{\se}$
and we found that the anisotropy parameter $A$
not only depends on the ratio of the ``thermal speeds'' $\Ts/\Tp$
but also on $\kappa$
and the cutoff parameters ($\ap,\as$).
Interestingly, the ratio $A$
can drop from values above one to those below one, and vice versa.
Also the heat flow vector for the RBK distribution shows a similar
interesting feature: The heat flow components as function of $\kappa$
can intersect for small values ($\kappa<1.5$).
The intersection point $\kappa_{s}$
depends on the components of the drift vector $\vec{W}$
(or macroscopic fluid vector).  These new features can have important
consequences for the interpretation of various properties of
anisotropic plasmas, for example dispersion and stability properties,
which will in future be studied in more detail.

\section*{Acknowledgements}
KS and HF are grateful to the
\emph{Deut\-sche For\-schungs\-ge\-mein\-schaft, DFG\/} funding the
projects SCHE334/10-1 and FI706/15-1, respectively.  ML acknowledges
support from the Katholieke Universiteit Leuven, Ruhr-University
Bochum and Alexander von Humboldt Foundation. These results were
obtained in the framework of the projects G0A2316N (FWO--Vlaanderen)
and SCHL~201/35-1 (DFG--German Research Foundation). We also
appreciate the support from the International Space Science Institute
(ISSI) for hosting the international ISSI team on Kappa Distributions:
From Observational Evidences via Controversial Predictions to a
Consistent Theory of Suprathermal Space Plasmas, which triggered many
fruitful discussions that were beneficial for the work presented here

\appendix
\section{Vector and tensor notation}\label{appVec}
The dyadic and higher products are given by:
\begin{eqnarray}
  \vv{v} \otimes \vv{v} &=& \tensor{V} = (v_{i}v_{j}) \\\nonumber
  \vv{v} \otimes \vv{v}\otimes \vv{v} &=& \tensor{VV} =
                                             (v_{i}v_{j}v_{k}) 
\end{eqnarray}
We have to distinguish between the heat flux tensor $\tensor{Q}$ and
the heat flux vector $\vv{q}$:
\begin{eqnarray}
  \tensor{Q} &=& \int \vv{v} \otimes \vv{v}\otimes \vv{v} f
  \mathrm{d}^{3}v\\
  \vv {q} &=& \int v^{2}\vv{v} \mathrm{d}^{3}v
\end{eqnarray}
The latter is also sometimes expressed as $\tr{Q}$.
\onecolumngrid
\subsection{Spherical coordinates}

We define the spherical velocity vector as
\begin{align}
  \vv{v} &= (v \cosf\sint,v \sinf\sint, v \cost)^{T}
\end{align}
so that, with $w =\frac{v}{\Theta}$
\begin{align}
   \Rightarrow \vv{v} = w\Theta(\cosf\sint,\sinf\sint, \cost)^{T}
\end{align}
and the volume element is
\begin{align}
  \mathrm{d}^{3}v = \Theta^{3}\vols
\end{align}
and the dyadic product  is:
\begin{align}\nonumber
  \vv{v}\otimes\vv{v} &= \tensor{V} = (v_{i}v_{j})=
v^{2}   \begin{pmatrix}
\cosf[2]\sint[2] & \cosf\sinf\sint[2] & \cosf\sint\cost\\
\cosf\sinf\sint[2]& \sinf[2]\sint[2]  & \sinf\sint\cost\\
\cosf\sint \cost   & \sinf\sint\cost     &  \cost[2]
\end{pmatrix}\\
  \vv{v}\otimes\vv{v}\otimes\vv{v} &
                                     = \tensor{VV} = (v_{i}v_{j}v_{k})
\end{align}
which is not explicitly given here, but it could
be easily written out.

\subsection{Cylindrical coordinates}
We define
\begin{align}
  \vv{v} &= (v\pa,v\se\cos\vartheta,v\se\sin\vartheta)^{T}
\end{align}
with
\begin{align}\nonumber
&w\pa   = \frac{v\pa}{\Theta\pa}, \qquad w\se  = \frac{v\se}{\Theta\se}\\
  &\Rightarrow \vv{v} = (\Theta\pa w\pa,\Theta\se w\se\cos\vartheta,\Theta\se w\se\sin\vartheta)^{T}
\end{align}
and the volume element is
\begin{align}
  \mathrm{d}^{3}v = \Theta\pa\Theta\se^{2}\volc
\end{align}
and the dyadic product  is:
\begin{align}
  \vv{v}\otimes\vv{v} = \tensor{V} = (v_{i}v_{j}) =
  \begin{pmatrix}
v\pa^{2} & v\pa v\se\cos{\vartheta}& v\pa v\se\sin{\vartheta}\\
v\pa v\se\cos{\vartheta} & v\se^{2}\cos^{2}{\vartheta}
&v\se^{2}\sin{\vartheta}\cos{\vartheta}\\
v\pa v\se\sin{\vartheta} & v\se^{2}\sin{\vartheta}\cos{\vartheta} &v\se^{2}\sin^{2}{\vartheta}  
\end{pmatrix}
\end{align}
\section{Moments and most probable parameter}\label{appMom}
In the following we drop the indices of the distribution functions $f$
and normalization constants $N$ to save writings. It is clear from
the context which of the above presented distribution functions is
meant in what follows. Note: All moments are normalised to the mass.

\subsection{Spherical coordinates}
In order to calculate the normalisation constant $N$, we use
$f^{\prime}= \frac{f}{n N }$ in the 0th order moment.
With the isotropic volume element the moments equations
Eq.~\ref{genmom} are
\begin{subequations}\label{spherepress}
  \begin{align}
  \frac{1}{N}& = 4\pi\Theta^{3}\int\limits_{0}^{\infty}f^{\prime} w^{2}\dw\\
u_{p} &= \frac{4\pi\Theta^{4}}{n} \int\limits_{0}^{\infty}w^{3}f\dw\\
  \tensor{P} &=  \Theta^{5}
               \bigintsss_{0}^{\infty}\bigintsss_{0}^{2\pi}\bigintsss_{0}^{\pi}
               \begin{pmatrix}
    \cosf[2]\sint[2] & 0 & 0\\
          0 & \sinf[2]\sint[2]  & 0\\\nonumber
0   & 0     &  \cost[2]
 \end{pmatrix}
              f   w^{4} \vls\\\label{eq:circ}
    &=  \Theta^{5}\frac{4}{3}\pi  \begin{pmatrix}
    1 & 0 & 0\\
          0 & 1  & 0\\
0   & 0     &  1
 \end{pmatrix}
          \int\limits_{0}^{\infty}    f   w^{4} \dw\\
  q_{p} &= 4\pi\Theta^{6} \int\limits_{0}^{\infty} w^{5}f \dw
  \end{align}
\end{subequations}

Only the quadratic terms in \sint\ and \cost\ survive for
$\tensor{P}$, and, thus, $P_{ij}=0, \ \ \forall\ i\ne j$.


\subsection{Cylindrical coordinates}
We introduce also the most probable speed
$u_{p\pa}$ along $w\pa$ and $u_{p\se}$ along $w\se$:

\begin{subequations}
\begin{align}\label{cyc:cor}
  \frac{1}{N}& =\Theta\pa\Theta\se^{2}\icyl f^{\prime} \volc\\\label{cyc:cor1}
u_{p} &= \frac{\Theta\pa \Theta^{2}\se}{n} \icyl
    \sqrt{w\pa^{2}\Theta\pa^{2}+w\se^{2} \Theta\se^{2}}w\se f \vo  \\\label{cyc:cor2}
u_{p\parallel}&= \frac{\Theta\pa^{2} \Theta^{2}\se}{n}
  \icyl
   w\pa w\se f \vo  \\
u_{p\perp}&= \frac{\Theta\pa \Theta^{3}\se}{n}
  \icyl
   w\se^{2} f \vo  \\\nonumber\label{cyc:cor3}
  \tensor{P}&= \Theta\pa \Theta^{2}\se \bigintsss_{0}^{2\pi}\bigintsss_{-\infty}^{{\infty}}\bigintsss_{0}^{\infty} \begin{pmatrix}
w\pa^{2}\Theta^{2}\pa & 0& 0\\
0 & w\se^{2}\Theta^{2}\se\cost[2]&0\\
0 & 0&w\se^{2}\Theta^{2}\se\sint[2]  
 \end{pmatrix}
       w\se f \vo\\  
             &=2\pi\Theta\pa \Theta^{2}\se \bigintsss_{-\infty}^{{\infty}} \bigintsss_{0}^{\infty}\begin{pmatrix}
w\pa^{2}\Theta^{2}\pa & 0& 0\\
0 & \frac{1}{2} w\se^{2}\Theta^{2}\se&0\\
0 & 0& \frac{1}{2}w\se^{2}\Theta^{2}\se  
 \end{pmatrix}
       w\se f \voc\\\nonumber  
  \\\label{cyc:cor4}
    q_{p} &= \Theta\pa \Theta\se^{2} \icyl
   (w\pa^{2}\Theta\pa^{2}+w\se^{2}\Theta\se^{2})^{3/2}w\se
 f \vo \\\label{cyc:cor5}
    q_{p\parallel} &= 2\Theta\pa^{4} \Theta\se^{2}
  \icyl w\pa^{3}w\se
 f\vo\\\label{cyc:cor6}
   q_{p\perp} &= 2\Theta\pa \Theta\se^{5}
  \icyl w\se^{4}
 f\vo\\
\end{align}
\end{subequations}

Unfortunately, the square root in the most probable parameter
Eq.~\ref{cyc:cor1}  and Eq.~\ref{cyc:cor4}  does
not allow, in general, for an analytic solution, as far as we know.
Nevertheless, the more interesting cases are the most probable speeds
along the parallel direction and in the perpendicular ones. The same
holds true for the most probable heat flux.


\subsection{With bulk speed $\vv{W}\ne \vv{0}$}
We replace $\vv{w}^{\prime}\rightarrow \vv{w}-\vv{W}$,
$\vv{w}^{\prime}\pa\rightarrow \vv{w}\pa-\vv{W}\pa$
and $\vv{w}^{\prime}\se\rightarrow \vv{w}\se-\vv{W}\se$,
respectively in the distribution functions (with
$W\pa\Theta\pa=U\pa$,
...).  We discuss here only the most probable speeds, the heat flux
vector, and the most probable heat flow:
\begin{subequations}\label{heatbulk}
\begin{align}
  u_{p}&= \frac{\Theta}{n_{0}}
  \int\limits_{0}^{\infty} \int\limits_{0}^{2\pi}\int\limits_{0}^{{\pi}}
   |\vv{w}+\vv{W}|^{3} f \vls  \\\label{heatbulk:1}
u_{p\parallel}&= \frac{2 \Theta\pa^{2} \Theta^{2}\se}{n_{0}}\icyl 
   |w\pa+W\pa| w\se f \vo  \\\label{heatbulk:2}
u_{p\perp}&= \frac{\Theta\pa \Theta^{3}\se}{n_{0}}
  \icyl
   (\vv{w}\se+\vv{W}\se)^{2} f \vo  \\\label{heatbulk:3}
  \vv{q}_{\mathrm{I}}&= \Theta^{6}\isph (2(\vv{w}\cdot\vv{W})\vv{w}+w^{2}\vv{W}+W^{2}\vv{W}) f_{}\vols\\\label{heatbulk:3a}
   \vv{q}_{\mathrm{A}}&= \Theta\pa\Theta^{2}\se\icyl (2(\vv{v}\cdot\vv{U})\vv{v}+v^{2}\vv{U}+U^{2}\vv{U}) f\volc\\\label{heatbulk:4}
  q_{p} &= \Theta^{6}\isph |\vv{w}+\vv{W}|^{5}f \vls\\\label{heatbulk:5}
    q_{p\parallel} &= \Theta\pa^{4} \Theta\se^{2}
     \icyl
     |w\pa+W\pa|^{3}w\se f_{\pi}\vo\\\label{heatbulk:6}
   q_{p\perp} &= \Theta\pa \Theta\se^{5}
  \icyl (\vv{w}\se+\vv{W}\se)^{4}
 f \vo
\end{align}  
\end{subequations}
For clarity, we used the following notation for the heat flow
vectors: $\vv{q}_{\mathrm{I}}$
for that of the isotropic distribution functions and
$\vv{q}_{\mathrm{A}}$
for that of the anisotropic ones, because they are slightly different.
Additionally, we have assumed that the modulus to power $2n$
is the same as the power $2n$:
$|\vv{a}\pm\vv{b}|^{2n}=(\vv{a}\pm\vv{b})^{2n}$.
For odd powers $2n+1$
we cannot decompose $|\vv{w}\pm\vv{W}|^{2n+1}$, which is the case for
 the spherical distribution functions. But for the cylindrical parallel speeds
$|w\pa\pm W\pa|^{2n+1}$ we can decompose in the integrals into
\begin{align}\label{decomp}
  |w\pa\pm W\pa|^{2n+1}=
  \begin{cases}
    ( w\pa + W\pa)^{2n+1} \qquad  &w\pa,W\pa >0\\
    ( w\pa - W\pa)^{2n+1} \qquad  &w\pa > W\pa\\
    (-w\pa + W\pa)^{2n+1} \qquad  &w\pa < W\pa\\
    (-w\pa - W\pa)^{2n+1} \qquad  &w\pa,W\pa <0\\
  \end{cases}
\end{align}

These cases can then be treated separately in the integral from
$-\infty$
to $\infty$,
see Appendix~\ref{appMost}, where care must be taken with the
integration boundaries.
For the perpendicular case in cylindrical coordinates we have always
power of $2n$ which can be decomposed.
We do not calculate  the heat flux tensor.


\subsubsection{Spherical coordinates}

The last integral of
Eq.~\ref{heatbulk:3} is the heat flow along $\vv{W}$ which yields
$ n\ W^{2}\vv{W}$.
The second integral
is the flow of the particle energy density along $\vv{W}$ resulting in
$\Theta P_{11}\vv{W}$ and the first is proportional to
\begin{align}
   2\Theta^{2}
\isph
          \begin{pmatrix}
    w W_{1}(W_{1}\cosf\sint + W_{2}\sinf\sint+ W_{3}v \cost)\\
    w W_{2}(W_{1}\cosf\sint + W_{2}\sinf\sint+ W_{3}v \cost)\\
    w W_{3}(W_{1}\cosf\sint + W_{2}\sinf\sint+ W_{3}v \cost)
  \end{pmatrix}
  f\vols = 0
\end{align}
and vanishes.  Thus, we have for the most probable heat flow:
\begin{align}\label{eq:heatdrift}
  \vv{q} = \Theta \left(P_{11} + n\Theta^{2}W^{2}\right)\vv{W} = n
  \Theta^{3} \vec{W}(\kappa \Ika[2][0](\kappa,\alpha) + W^{2})
\end{align}


\subsubsection{Cylindrical coordinates}
With $\vec{W} = (W\pa,W_{\se,1},W_{\se,2})^{T}$,
the first integral in Eq.~\ref{heatbulk:3a} gives
\begin{align}
  2\Theta\pa\Theta^{2}\se &\icyl
  (\vv{v}\cdot\vv{U})\vv{v}f_{}\volc =\\\nonumber
&  2\Theta\pa\Theta^{2}\se\icyl (\Theta\pa^{2} w\pa W\pa  +
  \Theta\se^{2}w\se W_{\se,1}\cost   + \Theta^{2}\se w\se W_{\se,2} \sint)
\begin{pmatrix}\Theta\pa w\pa \\ \Theta\se w\se\cost\\\Theta\se w\se\sint\end{pmatrix}
 f \volc = \\\nonumber
  &2\Theta\pa\Theta^{2}\se\
    \bigintsss_{0}^{\infty}\bigintsss_{-\infty}^{\infty}\
    \begin{pmatrix}
    2 \pi \Theta^{3}\pa w\pa^{2}w\se W\pa \\
   \  \pi \Theta^{3}\se w\se^{3} W_{\se,1}\\
    \ \pi \Theta^{3}\se w\se^{3} W_{\se,2}\\
  \end{pmatrix}
  f \mathrm{d}w\pa\mathrm{d}w\se\\\nonumber
 &\substack{=\\\mathrm{RBK}} \ \ n_{\mathrm{RBK}}\begin{pmatrix}
   \frac{3}{2}
   \Theta\pa^{3}\kappa^{\frac{3}{2}}\Jka[2,1][0,0] W\pa\\
    \frac{3}{4}
   \Theta\se^{3}\kappa\Jka[0,3][0,0] W_{\se,1}\\
    \frac{3}{4}
   \Theta\se^{3}\kappa\Jka[0,3][0,0] W_{\se,2}\\
 \end{pmatrix}
  = \kappa
  \begin{pmatrix}
   P_{11}  \Tp W\pa\\
   P_{22}  \Ts W_{\se,1}\\
   P_{22}   \Ts W_{\se,2}\\
 \end{pmatrix}
\end{align}
and similar with the RPBK distribution.
The last integral in Eq.~\ref{heatbulk:3a} gives $n W^{2} \vec{W}$. The
second results in
\begin{align}
   \Theta\pa\Theta^{2}\se
  \icyl
          (v\pa^{2}+v\se^{2}) f\volc =  ( \Theta\pa P_{11}+\Theta\se P_{22}).
\end{align}
and, thus, we have
\begin{align}\label{heataniso}
  \vec{q} = 2
  \begin{pmatrix}
   P_{11}  \Tp W\pa\\
   P_{22}  \Ts W_{\se,1}\\
   P_{22}   \Ts W_{\se,2}\\
 \end{pmatrix}
  + (  P_{11}+ P_{22})  \begin{pmatrix}
    \Theta\pa W\pa\\\Theta\se W_{\se,1} \\\Theta\se W_{\se,2}
   \end{pmatrix} 
 + [n  (\Theta\pa^{2} W^{2}\pa +
  \Theta^{2}\se W\se^{2})]
  \begin{pmatrix}
    \Tp W\pa\\\Ts W_{\se,1} \\ \Ts W_{\se,2}
   \end{pmatrix} 
\end{align}
%


\section{Solutions of integrals}\label{appInt}


\subsection{The integrals of regularized isotropic $\kappa$ distribution function}

The corresponding isotropic moments
$I(\kappa,\alpha,\nu,\Theta)\equiv\Ika[\nu](\kappa,\alpha)$
of the RK distribution are  \cite[see][]{Scherer-etal-2017}
\begin{subequations}
\begin{align}\label{eq:2}
  I(\kappa,\alpha,\nu,\Theta) &=
   \Theta^{3+\nu}
                \int\limits_{0}^{\infty}\left(1+\frac{w^{2}}{\kappa}\right)^{-\kappa-1}e^{-\alpha^{2}w^{2}}w^{2+\nu}\mathrm{d}w\\\nonumber
               &= \frac{1}{2}\Theta^{3+\nu}\kappa^{\frac{3+\nu}{2}}
                \int\limits_{0}^{\infty}\left(1+x\right)^{-\kappa-1}e^{-\alpha^{2}\kappa
                 x} x^{\frac{1+\nu}{2}}\mathrm{d}x\\\nonumber
            &=\frac{1}{2}
              \Theta^{3+\nu}\kappa^{\frac{3+\nu}{2}}\GA{\frac{3+\nu}{2}}
              U\left(\frac{3+\nu}{2},\frac{3-2\kappa+\nu}{2},\alpha^{2}\kappa\right)\\
  i(\kappa,\alpha,\nu,\Theta,W)&=
 \Theta^{3+\nu}\int\limits_{0}^{W}w^{\nu+2}  \left(1 + \frac{w^{2}}{\kappa
                               }\right)^{-\kappa-1} {\rm
                                   e}^{-\alpha^{2} w^{2}} \dw\\\nonumber
 &= \frac{1}{2}\Theta^{3+\nu}\kappa^{\frac{3+\nu}{2}}
                \int\limits_{0}^{W}\left(1+x\right)^{-\kappa-1}e^{-\alpha^{2}\kappa
                 x} x^{\frac{1+\nu}{2}}\mathrm{d}x
\end{align}
\end{subequations}
Where $U(a,b,z)$
is the Kummer~U or Tricomi function and $\Gamma$
is the Gamma-function , see \citet{Abramowitz-Stegun-1972},
\citet{Gradstein-Ryshik-1981}, or \citet{Oldham-etal-2010}. The above
representation of $I(\kappa,\alpha,\nu,\Theta)$
is more compact than that in \citet{Scherer-etal-2017} and was found
by \citet{Yoon-etal-2018}.

Only the radial part of the spherical volume element
$w^{2}\sint\dtheta\dphi\dw\rightarrow 4\pi w^{2}\dw$
was used, because the trigonometric part can easily be integrated.

To save writings, we define:
\begin{subequations}
\begin{align}\label{eq:Ikant} 
  \Ika[\nu][\eta](\kappa,\alpha) \equiv&
      \frac{U\left(\frac{3+\nu}{2},\frac{3-2\kappa+\nu}{2},\alpha^{2}\kappa\right)}
           {U\left(\frac{3+\eta}{2},\frac{3-2\kappa+\eta}{2},\alpha^{2}\kappa\right)}                               
  \\
  \ika[\nu][\eta] (\kappa,\alpha,W) \equiv&
                  \int\limits_{-\infty}^{W}\left(1+x\right)^{-\kappa-1}e^{-\alpha^{2}\kappa
                 x} x^{\frac{1+\nu}{2}}\mathrm{d}x
\end{align}
\end{subequations}
The first definition combines the $\kappa$ and $\alpha$ dependent parts
of normalisation with that from the moment $\nu$, which are
dimensionless. The dimensions of the corresponding moments are given by
powers of $\Theta$ and the number density
$n$. 

We have the following rules:
\begin{align}
  \Ika[][\eta] &\equiv \frac{1}{\Ika[\eta]}\\\nonumber
  \Ika[\nu][\eta] &\equiv \frac{\Ika[\nu]}{\Ika[\eta]}\\\nonumber
  \Ika[\nu][\eta]\Ika[\nu][\xi]&=\Ika[\nu][\xi]\\\nonumber
  \frac{\Ika[\nu][\eta]}{\Ika[\xi][\eta]}&=\Ika[\nu][\xi]\\\nonumber
  \Ika[\nu][\nu]&= 1
\end{align}
here are $\nu,\eta,\xi\in\mathbb{R}$.

So we find for the normalisation constant $N_{\mathrm{RK}}$
  (including a factor $2\pi$ from the volume element) and
for the elements of the moment tensors
$M^{\nu}_{\mathrm{RK}}$, including the number density
$n_{\mathrm{RK}}$.  We do not include the factors from the integration with
respect to the angle variable $\vartheta$ because they can differ with
the order of the moments. 
\begin{subequations}\label{mrk}
\begin{align}
  N_{\mathrm{RK}}&= \frac{1}{\Theta^{3}\sqrt{\pi^{3}\kappa^{3}}
                   }\Ika[][0]\\
  M^{\nu}_{\mathrm{RK}}
                 &=n_{\mathrm{RK}}N_{\mathrm{RK}}\frac{1}{2}\Theta^{3+\nu}\GA{\frac{3+\nu}{2}}\Ika[2][]
  \\\nonumber
  &=n_{\mathrm{RK}}\frac{2}{\sqrt{\pi}}\Theta^{\nu}\kappa^{\frac{\nu}{2}}\GA{\frac{3+\nu}{2}}\Ika[\nu][0]
\end{align}
\end{subequations}
To get the correct pressure the moment elements have to be multiplied
by $4\pi/3$ (see Eq.~\ref{spherepress}).

\subsection{The integrals for the product $f\pr$}
For our applications, we can assume that $\lambda$
  and $\mu$
  have integer values for the moments {\bf  of the distribution function} $f\pr$ (Eq.~\ref{eq:mod}) and, thus,
  the integral for odd values of $\lambda$ vanishes. In the case of the
most probable speed and heat flow we take twice the integral $\mathrm{d}w\pa$ from 0 to
infinity, assuming that $|w\pa|^{\lambda}=w\pa^{\lambda}$. With
this assumption, we find
\begin{subequations}
\begin{align}
    I(\kappa,&\ap,\lambda,\Theta\pa)I(\kappa,\as,\mu,\Theta\se)
  \\\nonumber
             &= \Theta\pa^{\lambda+1}\Theta\se^{\mu+2}
               \int\limits^{\infty}_{-\infty}\int\limits^{\infty}_{0}
               \left(1+ \frac{w\pa^{2}}{\kappa\pa
      }\right)^{-\kappa\pa-s\pa}\left(1+ \frac{w\se^{2}}{\kappa\se
      }\right)^{-\kappa\se-s\se}\mathrm{e}^{-\alpha\pa^{2}
      w\pa^{2}-\alpha\se^{2}  w\se^{2}}w\pa^{\lambda}w\se^{\mu+1} \mathrm{d}w\pa\mathrm{d}w\se\\
             &= 2\Theta\pa^{\lambda+1}\Theta\se^{\mu+2}
               \int\limits_{0}^{\infty}\left(1+ \frac{w\pa^{2}}{\kappa\pa
      }\right)^{-\kappa\pa-s\pa}  \mathrm{e}^{-\alpha\pa^{2}
    w\pa^{2}}w\pa^{\lambda}\mathrm{d}w\pa
    \int\limits_{0}^{\infty}\left(1+ \frac{w\se^{2}}{\kappa\se
      }\right)^{-\kappa\se-s\se}\mathrm{e}^{-\alpha\se^{2}
    w\se^{2}}w\se^{\mu+1}\mathrm{d}w\se\\
   &=\frac{1}{2}\Theta\pa^{\lambda+1}\Theta\se^{\mu+2}\kappa\pa^{\frac{\lambda+1}{2}}\kappa\se^{\frac{\mu+2}{2}}\left(\bigintsss \left(1+ x\right)^{-\kappa\pa-s\pa}  \mathrm{e}^{-\alpha\pa^{2}
    \kappa\pa x}x^{\frac{\lambda-1}{2}}\dx\right) \left(\bigintsss
    \left(1+ x\right)^{-\kappa\se-s\se}\mathrm{e}^{-\alpha\se^{2}
               \kappa\se x}x^{\frac{\mu}{2}}dx\right) \\\label{eq:pro}
  &=\frac{1}{2}\Theta\pa^{\lambda+1}\Theta\se^{\mu+2}\kappa\pa^{\frac{\lambda+1}{2}}\kappa\se^{\frac{\mu+2}{2}} \GA{\frac{\lambda+1}{2}}\GA{\frac{\mu+2}{2}}\Vka[\lambda][][\pa](\kappa\pa,\alpha\pa)\Vka[\mu][][\se](\kappa\se,\alpha\se)
\end{align}
\end{subequations}
where we have defined
\begin{align}\label{eq:tripro}
  \Vka[\lambda][][\pa](\kappa\pa,\alpha\pa) &\equiv
  U\left(\frac{\lambda+1}{2},\frac{\lambda+3}{2}-\kappa\pa-s\pa,\kappa\pa\alpha^{2}\pa\right)\\
    \Vka[\mu][][\se](\kappa\se,\alpha\se) &\equiv
  U\left(\frac{\mu+2}{2},\frac{\mu+4}{2}-\kappa\se-s\se,\kappa\se\alpha^{2}\se\right)
\end{align}
For $i\in\{\pa,\se\}$  we define analogously as above:
\begin{align}
  \Vka[x][y][i](\kappa_{i},\alpha_{i})   = 
   \frac{ \Vka[x][][i]}{ \Vka[][y][i]}(\kappa_{i},\alpha_{i})= \Vka[x][y][i]=\frac{ \Vka[x][][i]}{ \Vka[][y][i]}
\end{align}
Finally, we get for the normalization constant and the elements of the
moment tensors (see above Eqs.\ref{mrk} with a factor $2\pi$ from the
cylindrical volume element)
\begin{subequations}\label{eq:prodmom}
\begin{align}
  N_{\mathrm{RPBK}}^{\pa,\se}&= \frac{1}{\sqrt{\pi^{3}}\Theta\pa\Theta\se^{2}
                     \kappa_{\pa}^{\frac{1}{2}}\kappa\se}
                     \Vka[][0][\pa](\kappa\pa,\ap)\Vka[][0][\se](\kappa\se,\as)\\\nonumber
  M^{\lambda,\mu,s\pa,s\se}_{\mathrm{RPBK}}
                   &=n_{\mathrm{RPBK}}N_{\mathrm{RPBK}}^{s\pa,s\se}\frac{1}{2}
                     \Theta\pa^{\lambda+1}\Theta\se^{\mu+2}
                     \kappa^{\frac{\lambda+1}{2}}\pa
                     \kappa^{\frac{\mu+2}{2}}
    \GA{\frac{\lambda+1}{2}}\GA{\frac{\mu+2}{2}}
                     \Vka[\lambda][][\pa](\kappa\pa,\ap)\Vka[\mu][][\se](\kappa\se\as)
    \,\Pi
  \\\label{mrpbk}
  &=n_{\mathrm{RPBK}}\frac{1}{2\sqrt{\pi^{3}}}
    \Theta\pa^{\lambda}\Theta\se^{\mu}
    \kappa\pa^{\frac{\lambda}{2}}\kappa\se^{\frac{\mu}{2}}
    \GA{\frac{\lambda+1}{2}}\GA{\frac{\mu+2}{2}}
    \Vka[\lambda][0][\pa](\kappa\pa,\ap)\Vka[\mu][0][\se](\kappa\se,\as)
    \,\Pi
\end{align}
\end{subequations}
To get the correct pressure elements ($P_{11},P_{22}=P_{33}$)
we have to multiply the above moment elements by the factors
$\Pi$ as calculated in the matrix equation Eq.~\ref{cyc:cor3}.
 \begin{align}\label{eq:pi}
   \Pi =
  \begin{cases}
    2\pi \qquad \mathrm{for}\ P_{11}\\  
    \phantom{2}\pi \qquad \mathrm{for}\ P_{22}\ \mathrm{and}\ P_{33}\\  
    \end{cases}
\end{align}
Similar scaling factors have to be calculated for higher-order
moments, which are not discussed here. For the most probable speeds
and heat flows a factor $2\pi$ as in the normalisation constant
$N_{\mathrm{RPBK}}$ needs to be multiplied.

  %

\subsection{The moments of $f\su$}
The  $f\su$ is given by
\begin{align}
   f\su  &=
         f\su(\kappa,\alpha\pa,\alpha\se,\Theta\pa,\Theta\se,v\pa,v\se) \\\nonumber
   &= n_{0}N\su
      \left(1+ \frac{v\pa^{2}}{\kappa
      \Theta\pa^{2}} + \frac{v\se^{2}}{\kappa
      \Theta\se^{2}}\right)^{-\kappa-1}\mathrm{e}^{-\frac{\alpha\pa^{2}
      v\pa^{2}}{\Theta\pa^{2}}-\frac{\alpha\se^{2}
      v\se^{2}}{\Theta\se^{2}}}
\end{align}
Thus, we have  to solve the integrals below,  where we do not take
  into account the integration with respect to $\vartheta$ which is
  straight forward integration using the equations from Appendix~A. To save writings we
define: $J(\kappa,\alpha\pa,\alpha\se,\lambda,\mu,\Theta\pa,\Theta\se)\equiv J$
and  $
j^{\pa}(\kappa,\alpha\pa,\alpha\se,\lambda,\mu,\Theta\pa,\Theta\se)\equiv
j^{\pa}$ and, analogously, for the perpendicular direction. 
\begin{align}\label{Jka}
  J &= \Theta_{\pa}^{1+\lambda}\Theta_{\se}^{2+\mu} \int\limits_{-\infty}^{\infty}\int\limits_{0}^{\infty}
   w\pa^{\lambda}w\se^{\mu+1}\left(1+ \frac{w\pa^{2}}{\kappa} +
  \frac{w\se^{2}}{\kappa}\right)^{-\kappa-1}\mathrm{e}^{-\alpha\pa^{2}w\pa^{2}-\alpha\se^{2}
  w\se^{2}} \dw\se\dw\pa\\
   j^{\pa} &= \Theta_{\pa}^{1+\lambda}\Theta_{\se}^{2+\mu} \int\limits_{-\infty}^{W\pa}\int\limits_{0}^{\infty}
   w\pa^{\lambda}w\se^{\mu+1}\left(1+ \frac{w\pa^{2}}{\kappa} +
  \frac{w\se^{2}}{\kappa}\right)^{-\kappa-1}\mathrm{e}^{-\alpha\pa^{2}w\pa^{2}-\alpha\se^{2}
  w\se^{2}} \dw\se\dw\pa\\
     j^{\se} &= \Theta_{\pa}^{1+\lambda}\Theta_{\se}^{2+\mu}\int\limits_{-\infty}^{\infty}\int\limits_{0}^{W\se}
   w\pa^{\lambda}w\se^{\mu+1}\left(1+ \frac{w\pa^{2}}{\kappa} + \frac{w\se^{2}}{\kappa}\right)^{-\kappa-1}\mathrm{e}^{-\alpha\pa^{2}w\pa^{2}-\alpha\se^{2} w\se^{2}} \dw\se\dw\pa
\end{align}
  With the following substitution:
\begin{align}
  w\se = r \sint \qquad w\pa = r \cost, \qquad r = \sqrt{w\se^{2}+w\pa^{2}} 
\end{align}

\begin{align}\nonumber
J &= \Theta_{\pa}^{1+\lambda}\Theta_{\se}^{2+\mu} \int\limits_{0}^{\pi}\int\limits_{0}^{\infty}
r^{\lambda+\mu+2} \left(1+\frac{r^{2}}{\kappa }\right)^{-\kappa-1}
             \mathrm{e}^{-r^{2}(\as^{2}+(\ap^{2}-\as^{2})\cost[2])}\sint[\mu+1]\cost[\lambda]\dr\dtheta\\\nonumber
\end{align}
 The integral with respect to $r$
is similar to $I(\kappa,\alpha,\nu,\Theta)$.
Thus, we have to solve the following type of integral  (with
$a^{2}=\ap^{2}-\as^{2}$), which we take twice from $0$ to
$\frac{\pi}{2}$ to account for the cases when we want to calculate the
most probable parameters:
\begin{align} \label{int1}
   J &= \Theta^{1+\lambda}_{\pa}\Theta^{2+\mu}_{\se}\kappa^{\frac{3+\lambda+\mu}{2}}
                \int\limits_{0}^{\frac{\pi}{2}}\int\limits_{0}^{\infty}
                \left(1+x\right)^{-\kappa-1}e^{-\as^{2}\kappa x}
                e^{-\kappa a^{2} x \cost[2]}x^{\frac{1+\lambda+\mu}{2}}
                \sint[\mu+1]\cost[\lambda]
                \mathrm{d}x\dtheta\\\nonumber
  &=\Theta^{1+\lambda}_{\pa}\Theta^{2+\mu}_{\se}\kappa^{\frac{3+\lambda+\mu}{2}}
                \int\limits_{0}^{1}\int\limits_{0}^{\infty}
                \left(1+x\right)^{-\kappa-1}e^{-\as^{2}\kappa x}
                e^{-\kappa a^{2} x t^{2}}x^{\frac{1+\lambda+\mu}{2}}
                (1-t^{2})^{\frac{\mu}{2}}t^{\lambda}
    \mathrm{d}x\dt\\
  &=
    \Theta^{1+\lambda}_{\pa}\Theta^{2+\mu}_{\se}\kappa^{\frac{3+\lambda+\mu}{2}}
    \GA{\frac{3+\lambda+\mu}{2}}\\\nonumber
&\hspace*{1cm}    \int\limits_{0}^{1}
    U\left(\frac{3+\lambda+\mu}{2},\frac{3+\lambda+\mu}{2}-\kappa,
    \kappa\left[\as^{2}+(\ap^{2}-\as^{2}) t^{2}\right]\right) (1-t^{2})^{\frac{\mu}{2}}t^{\lambda}\dt 
\end{align}
The integral cannot be solved in general and a numeric solution is
required. Nevertheless, we can define 
\begin{align}\label{numer}
 & \Jka[\lambda,\mu](\kappa,\ap,\as)\equiv
  \int\limits_{0}^{1}U\left(\frac{3+\lambda+\mu}{2},\frac{3+\lambda+\mu}{2}-\kappa,\kappa\left[\as^{2}+(\ap^{2}-\as^{2})
   t^{2}\right]\right)(1-t^{2})^{\frac{\mu}{2}}t^{\lambda}\dt\\
   & \jka[\lambda,\mu]^{\pa}(\kappa,\ap,\as,W\pa)\equiv \int\limits_{-\infty}^{W\pa}\int\limits_{0}^{\infty}
   w\pa^{\lambda}w\se^{\mu+1}\left(1+ \frac{w\pa^{2}}{\kappa} +
  \frac{w\se^{2}}{\kappa}\right)^{-\kappa-1}\mathrm{e}^{-\alpha\pa^{2}w\pa^{2}-\alpha\se^{2}
  w\se^{2}} \dw\se\dw\pa\\
   & \jka[\lambda,\mu]^{\se}(\kappa,\ap,\as,W\se)\equiv \int\limits_{-\infty}^{W\se}\int\limits_{0}^{\infty}
   w\pa^{\lambda}w\se^{\mu+1}\left(1+ \frac{w\pa^{2}}{\kappa} +
  \frac{w\se^{2}}{\kappa}\right)^{-\kappa-1}\mathrm{e}^{-\alpha\pa^{2}w\pa^{2}-\alpha\se^{2}
  w\se^{2}} \dw\se\dw\pa
\end{align}
and solve the remaining integral numerically.  
If $\ap=\as\equiv\alpha$ the above integral reduces to 
\begin{align}\label{Uequ}
  \Jka[\lambda,\alpha](\kappa,\alpha,\alpha) &=\frac{1}{2}
                \frac{\GA{\frac{1+\lambda}{2}}\GA{\frac{2+\mu}{2}}}
                      {\GA{\frac{3+\lambda+\mu}{2}}}
                  \Ika[\lambda+\mu](\kappa,\alpha)
\end{align}
The above solution still depends on the perpendicular and parallel
values $\Theta\pa$ and $\Theta\se$ and the power indices $\lambda,\mu$
in the $\Gamma$ functions:
 $\GA{\frac{1+\lambda}{2}}$ and
$\GA{\frac{2+\mu}{2}}$. 

 We have the following identities:
\begin{align}
  \Jka[\lambda,\mu] &= \Jka[][\lambda,\mu]^{-1}\\
  \Jka[\lambda,\mu][\nu,\eta] &\equiv
  \Jka[\lambda,\mu]\Jka[][\nu,\eta]\\
  \Jka[\lambda,\mu][\lambda,\mu] &= 1\\
   \Jka[\lambda,\mu] &\ne  \Jka[\mu,\lambda]
\end{align}
Again we find for the normalization and tensor elements:
\begin{itemize}
\item for $\ap\ne\as$ 
\begin{subequations}\label{mrpbk}
\begin{align}
  N_{\mathrm{RBK}}&= \frac{1}{\sqrt{\pi^{3}}\Theta\pa\Theta\se^{2}\sqrt{\kappa^{3}}  } \Jka[][0,0](\kappa,\as)\\
  M^{\lambda,\mu}_{\mathrm{RBK}}
                  &=n_{\mathrm{RBK}}N_{\mathrm{RBK}}\,\Theta^{\lambda+1}\pa
                    \Theta\se^{\mu+2}\kappa^{\frac{\lambda+\mu+3}{2}}\GA{\frac{3+\lambda+\mu}{2}}
                    \Jka[\lambda,\mu](\kappa,\ap,\as)
                    \,\Pi
  \\\nonumber
                  &=n_{\mathrm{RBK}}\,\frac{1}{\sqrt{\pi^{3}}}\,\Theta^{\lambda}\pa\Theta\se^{\mu}
                    \kappa^{\frac{\lambda+\mu}{2}}
                    \GA{\frac{3+\lambda+\mu}{2}}\Jka[\lambda,\mu][0,0](\kappa,\ap,\as)
                    \,\Pi
\end{align}
\end{subequations}
\item for $\ap=\as=\alpha$
  \begin{subequations}\label{mrpbke}
\begin{align}
  N_{\mathrm{RPK}}&= \frac{1}{\sqrt{\pi^{3}}\Theta\pa\Theta\se^{2}\sqrt{\kappa^{3}}} \Ika[][0](\kappa,\alpha)\\
  M^{\lambda,\mu}_{\mathrm{RBK}}
                  &=n_{\mathrm{RBK}}N_{\mathrm{RBK}}\,\Theta^{\lambda+1}\pa
                    \Theta\se^{\mu+2}\kappa^{\frac{\lambda+\mu+3}{2}}
                    \GA{\frac{\lambda+1}{1}}\GA{\frac{\mu+2}{2}} \Ika[\lambda+\mu]
                    \,\Pi
  \\\nonumber
                  &=n_{\mathrm{RBK}}\,\frac{1}{\sqrt{\pi^{3}}}\,\Theta^{\lambda}\pa\Theta\se^{\mu}
                    \kappa^{\frac{\lambda+\mu}{2}}
                    \GA{\frac{\lambda+1}{2}}\GA{\frac{\mu+2}{2}}
                    \Ika[\lambda+\mu][0](\kappa,\alpha)
                    \,\Pi
\end{align}
\end{subequations}
\end{itemize}
and for the pressure elements $\Pi$ is given Eq.~\ref{eq:pi}.

\subsection{The moments of the distribution functions $f_{K},f_{BM}$, and $f_{PBM}$}

The moments of the distribution functions $f_{K}$, and
$f_{\mathrm{BK}}$ can be obtained setting $\alpha=\ap=\as$ in Eq.~\ref{eq:2}
and Eq.~\ref{Uequ}, where care must be taken, because the second
argument of the Tricomi function  should be lower than one:
\begin{align}\label{eq:lim}
  &\lim_{\substack{
\alpha^{2}\kappa \rightarrow0\\
\frac{3+\nu}{2}-\kappa<1}}
  U\left(\frac{3+\nu}{2},\frac{3+\nu}{2}-\kappa,\alpha^{2}\kappa\right)
  = \frac{\GA{\kappa-\frac{1+\nu}{2}}}{\GA{\kappa+1}}\\\nonumber
 &\lim_{\substack{
\alpha\pa^{2}\kappa\pa \rightarrow0\\
\frac{3+\lambda}{2}-\kappa-s\pa<1}} \Vka[\lambda][][s\pa] = \lim_{\substack{
\alpha\pa^{2}\kappa\pa \rightarrow0\\
\frac{3+\lambda}{2}-\kappa\pa-s\pa<1}}
  U(\left(\frac{1+\lambda}{2},\frac{3+\lambda}{2}-\kappa\pa-s\pa \right)
  =\frac{\GA{\kappa\pa+s\pa-\frac{\lambda+1}{2}}}{\GA{\kappa\pa+s\se}}\\\nonumber
 &\lim_{\substack{
\alpha\se^{2}\kappa\se \rightarrow0\\
\frac{\mu+4}{2}-\kappa\se-s\se<1}} \Vka[\mu][][s\se] = \lim_{\substack{
\alpha\se^{2}\kappa\pa \rightarrow0\\
\frac{\mu+4}{2}-\kappa\se-s\se<1}}
  U(\left(\frac{\mu+2}{2},\frac{\mu+4}{2}-\kappa\se-s\se \right)
  =\frac{\GA{\kappa\se+s\se-\frac{\mu+2}{2}}}{\GA{\kappa\se+s\se}}
\end{align}
Note, the factors for the integration over the angle variable are
  only included in the normalisation, and have to be handled for the
  tensor elements as above.
Thus, the moments for $f_{K}$ are:
\begin{align}\label{eq:K}
  N_{K} &= \frac{2\GA{\kappa+1}}{\Theta^{3}\sqrt{\pi^{3}\kappa^{3}}\GA{\kappa-\frac{1}{2}}}\\\nonumber
  M_{K}^{(\nu)}&=
                 N_{K}\Theta^{\nu+3}\kappa^{\frac{3+\nu}{2}}\GA{\frac{3+\nu}{2}}\frac{\GA{\kappa-\frac{1+\nu}{2}}}{\GA{\kappa+1}}\\
                 &= \frac{1}{\sqrt{\pi}} \Theta^{\nu}\kappa^{\frac{\nu}{2}} \GA{\frac{3+\nu}{2}}\frac{\GA{\kappa-\frac{1+\nu}{2}}}{\GA{\kappa-\frac{1}{2}}}
\end{align}
and those for $f_{BK}$:
\begin{align}\label{eq:BK}
  N_{BK} &=
                \frac{\GA{\kappa+1}}{\sqrt{\pi^{3}\kappa^{3}}\Theta_{\pa}\Theta^{2}_{\se}\GA{\kappa-\frac{1}{2}}}\\
   M_{BK}^{(\lambda,\mu)}&= \frac{N_{BK}}{\sqrt{\pi^{3}}}
                           \Theta_{\pa}^{1+\lambda}\Theta_{\se}^{2+\mu}\kappa^{\frac{3+\lambda+\mu}{2}}
                           \frac{\GA{\frac{1+\lambda}{2}}\GA{\frac{2+\mu}{2}}\GA{\kappa-\frac{1+\lambda+\mu}{2}}}{\GA{\kappa+1}}\,\Pi\\
  &=
    \frac{1}{2\sqrt{\pi^{3}}}\Theta_{\pa}^{\lambda}\Theta_{\se}^{\mu}\kappa^{\frac{\lambda+\mu}{2}}
    \frac{\GA{\frac{1+\lambda}{2}}\GA{\frac{2+\mu}{2}}\GA{\kappa-\frac{1+\lambda+\mu}{2}}}{\GA{\kappa-\frac{1}{2}}}\,\Pi
\end{align}
If $\Theta\pa=\Theta\se$ we get the values for $N_{K}$ and
  $M_{K}^{(\nu)}$ with  $\lambda+\mu=\nu$, except for
  $\GA{\frac{1+\lambda}{2}}\GA{\frac{2+\mu}{2}} \ne
  \GA{\frac{3+\nu}{2}}$, but $\lambda=\nu+2$ and $\mu=0$.

For the $f_{PBK}$ distribution we must be a little more careful,
because of the product, but nevertheless, we can use the limiting
approach from Eq.~\ref{eq:pro}:
\begin{align}\label{eq:PBK}
  N_{\mathrm{PBK}} & = \frac{\GA{\kappa\pa+s\pa}\GA{\kappa\se+s\se}}
                     {\sqrt{\pi^{3}}\sqrt{\kappa_{\pa}}\kappa\se
                     \Theta_{\pa}\Theta^{2}_{\se}
                     \GA{\kappa_{\pa}+s\pa-\frac{1}{2}}\GA{\kappa\se+s\se-1}}  \\\nonumber
  M_{\mathrm{PBK}}^{(\lambda,\mu)} &= n_{\mathrm{PBK}}N_{\mathrm{PBK}}
        \frac{1}{2}\Theta\pa^{\lambda+1}\Theta\se^{\mu+2}\kappa\pa^{\frac{\lambda+1}{2}}\kappa\se^{\frac{\mu+2}{2}}
      \GA{\frac{\lambda+1}{2}}\GA{\frac{\mu+2}{2}}
      \frac{\GA{\kappa_{\pa}+s\pa-\frac{\lambda+1}{2}}}{\GA{\kappa_{\pa}+s\pa}} 
                            \frac{\GA{\kappa_{\se}+s\se-\frac{\mu+2}{2}}}{\GA{\kappa_{\se}+s\se}}\,\Pi\\\label{eq:pbk}
          &= n_{\mathrm{PBK}} \frac{\Theta_{\pa}^{\lambda}\Theta_{\se}^{\mu}}{2\sqrt{\pi^{3}}}
            \kappa_{\pa}^{\frac{\lambda}{2}}\kappa_{\se}^{\frac{\mu}{2}}\,
            \GA{\frac{\lambda+1}{2}}\GA{\frac{\mu+2}{2}}
            \frac{\GA{\kappa_{\pa}+s\pa-\frac{\lambda+1}{2}}\GA{\kappa_{\se}+s\se-\frac{\mu+2}{2}}}{\GA{\kappa_{\pa}+s\pa-\frac{1}{2}}\GA{\kappa_{\se}+s\se-1}}\,\Pi  
\end{align}
In the limit $\Theta\pa=\Theta\se$ and $\kappa\pa=\kappa\se$ the
  isotropic case seems to be completely different, because of the
  different $\Gamma$-functions in Eq.~\ref{eq:pbk}.

The moments for the isotropic case are found in
Table~\ref{tab:02a} and that for the anisotropic distributions
in Table~\ref{tab:02}.

The condition $\frac{3+\nu}{2}-\kappa<1$ is reflected in those for
the standard $\kappa$-distributions, whose moments diverge below
those values. 
\section{The most probable parameters, $u_{p},u_{p\pa},u_{p\se},q_{p},q_{p\pa},q_{p\se}$}
\label{appMost}

\subsection{With: $W\pa=0,\vv{W}\se=\vv{0}$}
We process the integrals as follows (with
$\tilde{f}(w\pa,w\se)\equiv f(w\pa,w\se,\ap,\as)/(n N)$ and
dropping the indices):
\begin{align}\nonumber
\frac{1}{\Tp^{2}\Ts^{2}N}  u_{p\pa}&(W\pa=0)=\int\limits_{0}^{2\pi}
                \int\limits_{0}^{\infty}\int\limits_{-\infty}^{\infty}|w\pa|
                \tilde{f}(w\pa,w\se)\volc \\\nonumber
&=\int\limits_{0}^{2\pi}
                \int\limits_{0}^{\infty}\int\limits_{-\infty}^{0}|w\pa|
   \tilde{f}(w\pa,w\se)\volc +\int\limits_{0}^{2\pi}
                \int\limits_{0}^{\infty}\int\limits_{0}^{\infty}w\pa
                \tilde{f}(w\pa,w\se)\volc \\\nonumber
&=\int\limits_{0}^{2\pi}
                \int\limits_{0}^{\infty}\int\limits_{\infty}^{0}|-w\pa|
   \tilde{f}(-w\pa,w\se) w\se\mathrm{d}(-w)\pa\mathrm{d}w\se\mathrm{d}\vartheta+\int\limits_{0}^{2\pi}
                \int\limits_{0}^{\infty}\int\limits_{0}^{\infty}w\pa
                                                      \tilde{f}(w\pa,w\se)\volc \\\nonumber
  &=2\int\limits_{0}^{2\pi}
                \int\limits_{0}^{\infty}\int\limits_{0}^{\infty}w\pa
                                                      \tilde{f}(w\pa,w\se)\volc
\end{align}
where we have substituted in the first integral of the third line
$w\pa\rightarrow-w\pa$ and then changing the integral boundaries leads
to the fourth line, because $f(-w\pa,w\se)=f(w\pa,w\se)$. 
For the most probable heat flow $q_{p,\pi\pa}$ we find analogously: 
\begin{align}\nonumber
\frac{1}{n \Tp^{4}\Ts^{2}N}  q_{p\pa}&(W\pa=0)=\int\limits_{0}^{2\pi}
                \int\limits_{0}^{\infty}\int\limits_{-\infty}^{\infty}|w\pa|^{3}
                \tilde{f}(w\pa,w\se)\volc
   =2\int\limits_{0}^{2\pi}
                \int\limits_{0}^{\infty}\int\limits_{0}^{\infty}w\pa^{3}
    \tilde{f}(w\pa,w\se)\volc
\end{align}


\subsection{With $W\pa\ne0,\vv{W}\se=\vv{0}$}

\subsubsection{The most probable speed}
Now with a positive shift $W\pa>0$ in the distribution function
\begin{align}\label{f1}
\frac{1}{\Tp^{2}\Ts^{2}N} u_{p\pa}&(W\pa\ne0)=\int\limits_{0}^{2\pi}
                \int\limits_{0}^{\infty}\int\limits_{-\infty}^{\infty}|w\pa|
               \tilde{f}(w\pa-W\pa,w\se)\volc \\\nonumber
  &=\int\limits_{0}^{2\pi}
                \int\limits_{-\infty}^{\infty}\int\limits_{-\infty}^{0}|w\pa|
   \tilde{f}(w\pa^{\prime},w\se)\volc  \\\nonumber
&w^{\prime}\pa = w\pa + W\pa \qquad \mathrm{and} \qquad w\pa^{\prime} = w\pa-W\pa\\
 &=\int\limits_{0}^{2\pi}
                \int\limits_{0}^{\infty}\int\limits_{-\infty}^{\infty}|w^{\prime}\pa+W\pa|
                \tilde{f}(w\pa^{\prime},w\se)w\se\mathrm{d}w\pa^{\prime}\mathrm{d}w\se\mathrm{d}\vartheta
  \\\nonumber
  &=\int\limits_{0}^{2\pi}
                \int\limits_{0}^{\infty}\int\limits_{-\infty}^{0}|w^{\prime}\pa+W\pa|
                \tilde{f}(w\pa^{\prime},w\se)w\se\mathrm{d}w\pa^{\prime}\mathrm{d}w\se\mathrm{d}\vartheta
    +\int\limits_{0}^{2\pi}
                \int\limits_{0}^{\infty}\int\limits_{0}^{\infty}|w^{\prime}\pa+W\pa|
                \tilde{f}(w\pa^{\prime},w\se)w\se\mathrm{d}w\pa^{\prime}\mathrm{d}w\se\mathrm{d}\vartheta 
  \\\nonumber 
\end{align}
Now we replace $w\pa^{\prime}=-w\pa$
and replace the modulus according to the decomposition Eq.~\ref{decomp} by
$|W\pa-\tilde{w}\pa| \rightarrow (W\pa-\tilde{w}\pa)$
for $ W\pa>\tilde{w}\pa$
and $|W\pa-\tilde{w}\pa| \rightarrow (\tilde{w}\pa-W\pa)$
for $W\pa<\tilde{w}\pa$  in the first integral, while the last is
expressed as $u_{\pi,p\pa}(0)$ and $W\pa$
\begin{align}\label{eq:mpswp}\nonumber
  u_{p\pa}&= \Tp^{2}\Ts^{2}N \int\limits_{0}^{2\pi}
                \int\limits_{0}^{\infty}\int\limits_{W\pa}^{\infty}(w\pa-W\pa)
    \tilde{f}(w\pa,w\se)w\se\mathrm{d}w\pa\mathrm{d}w\se\mathrm{d}\vartheta \\\nonumber
&\hspace*{1cm}+\Tp^{2}\Ts^{2}N\int\limits_{0}^{2\pi}
                \int\limits_{0}^{\infty}\int\limits_{0}^{W\pa}(W\pa-w\pa)
                \tilde{f}(w\pa,w\se)w\se\mathrm{d}w\pa\mathrm{d}w\se\mathrm{d}\vartheta +
    \frac{1}{2}u_{\pi,p\pa}(0) +
    \frac{1}{2}\Theta\pa W\pa\\\nonumber
  &= \Tp^{2}\Ts^{2}N \int\limits_{0}^{2\pi}
                \int\limits_{0}^{\infty}\int\limits_{0}^{W\pa}(w\pa-W\pa)
    \tilde{f}(w\pa,w\se)w\se\mathrm{d}w\pa\mathrm{d}w\se\mathrm{d}\vartheta \\\nonumber
&\hspace*{1cm}-\Tp^{2}\Ts^{2}N\int\limits_{0}^{2\pi}
                \int\limits_{0}^{\infty}\int\limits_{0}^{W\pa}(W\pa-w\pa)
                \tilde{f}(w\pa,w\se)w\se\mathrm{d}w\pa\mathrm{d}w\se\mathrm{d}\vartheta +\nonumber
    u_{\pi,p\pa}(0)\\
   & = 2\Tp^{2}\Ts^{2}N\int\limits_{0}^{2\pi}
                \int\limits_{0}^{\infty}\int\limits_{0}^{W\pa}(W\pa-w\pa)
    \tilde{f}(w\pa,w\se)w\se\mathrm{d}w\pa\mathrm{d}w\se\mathrm{d}\vartheta 
   + u_{\pi,p\pa}(0) 
\end{align}
When $W\pa$ is close to zero, the integral vanishes and we are left
with $u_{p\pa}(0)$, while when $W\pa$ becomes large,  we
can assume that the term $u_{\pi,p\pa}(0)$ cancels with that in
the integral, and we are left with $W\pa\Theta\pa$.


\subsubsection{The most probable heat flow}
We proceed again analogously to Eq.~\ref{f1} (assuming $W\pa<0$) and find
\begin{align}\label{f2}
\frac{1}{n \Tp^{4}\Ts^{2}N} q_{p\pa}&(W\pa\ne0)=\int\limits_{0}^{2\pi}
                \int\limits_{0}^{\infty}\int\limits_{-\infty}^{\infty}|w\pa|^{3}
               \tilde{f}(w\pa-W\pa,w\se)\volc \\\nonumber
  &=\int\limits_{0}^{2\pi}
                \int\limits_{0}^{\infty}\int\limits_{-\infty}^{0}|w^{\prime}\pa+W\pa|^{3}
    \tilde{f}(w\pa^{\prime},w\se)w\se\mathrm{d}w\pa^{\prime}\mathrm{d}w\se\mathrm{d}\vartheta
    +\int\limits_{0}^{2\pi}
                \int\limits_{0}^{\infty}\int\limits_{0}^{\infty}|w^{\prime}\pa+W\pa|^{3}
                \tilde{f}(w\pa^{\prime},w\se)w\se\mathrm{d}w\pa^{\prime}\mathrm{d}w\se\mathrm{d}\vartheta 
  \\\nonumber
  & = \mathcal{I}_{1} +  \mathcal{I}_{2} 
\end{align}
We replace $|w^{\prime}\pa+W\pa|^{3}$ by $(w^{\prime}\pa+W\pa)^{3}$ in
the second integral $\mathcal{I}_{2}$, and in the first integral $\mathcal{I}_{1}$ we apply the
decomposition~\ref{decomp}
\begin{align}
  |w^{\prime}\pa+W\pa|^{3} = 
  \begin{cases}
(W\pa-w^{\prime}\pa)^{3} & W\pa > w\pa^{\prime}\\    
(w^{\prime}\pa-W\pa)^{3} & W\pa < w\pa^{\prime}\\    
  \end{cases}
\end{align}
Then the second integral $\mathcal{I}_{2}$ gives (neglecting the primes):
\begin{align}
n \Tp^{4}\Ts^{2}N \mathcal{I}_{2}&= n \Tp^{4}\Ts^{2}N\int\limits_{0}^{2\pi}
                \int\limits_{0}^{\infty}\int\limits_{0}^{\infty}(w\pa+W\pa)^{3}
                  \tilde{f}(w\pa,w\se)w\se\mathrm{d}w\pa\mathrm{d}w\se\mathrm{d}\vartheta\\\nonumber
  &=n \Tp^{4}\Ts^{2}N\int\limits_{0}^{2\pi}
    \int\limits_{0}^{\infty}\int\limits_{0}^{\infty}
    (w\pa^{3}+ 3w\pa^{2} W\pa + 3 w\pa W\pa^{2} + W\pa^{3})
    \tilde{f}(w\pa,w\se)w\se\mathrm{d}w\pa\mathrm{d}w\se\mathrm{d}\vartheta\\\nonumber
  &= \frac{1}{2}q_{p,\Pi\pa}(0) + \frac{3}{2}\Theta\pa W\pa P_{11,\Pi}(0)  + \frac{3}{2}
    \Theta\pa^{2 } W\pa^{2}u_{p,\Pi\pa}(0) +\pi n\Tp^{4}\Ts^{2} N\pa W\pa^{3} \int\limits_{0}^{\infty}\int\limits_{0}^{\infty}\tilde{f}(w\pa,w\se)\mathrm{d}w\pa\mathrm{d}w\se\mathrm{d}\vartheta
\end{align}

The first integral  $\mathcal{I}_{1}$  gives (changing $w^{\prime}\rightarrow-w$):
\begin{align}
  n \Tp^{4}\Ts^{2}N \mathcal{I}_{2}&= n \Tp^{4}\Ts^{2}N \int\limits_{0}^{2\pi}
                \int\limits_{0}^{\infty}\int\limits_{0}^{\infty}|-w\pa+W\pa|^{3}
    \tilde{f}(w\pa,w\se)w\se\mathrm{d}w\pa\mathrm{d}w\se\mathrm{d}\vartheta
  \\
  &= n\Tp^{4}\Ts^{2} N \int\limits_{0}^{2\pi}
                \int\limits_{0}^{\infty}\int\limits_{0}^{\infty}|-w\pa+W\pa|^{3}
    \tilde{f}(w\pa,w\se)w\se\mathrm{d}w\pa\mathrm{d}w\se\mathrm{d}\vartheta\\
  &=n \Tp^{4}\Ts^{2}N \int\limits_{0}^{2\pi}
                \int\limits_{0}^{\infty}\int\limits_{0}^{W\pa}(W\pa-w\pa)^{3}
    \tilde{f}(w\pa,w\se)w\se\mathrm{d}w\pa\mathrm{d}w\se\mathrm{d}\vartheta
  \\\nonumber
  &\hspace{1cm}+ n \Tp^{4}\Ts^{2}N \int\limits_{0}^{2\pi}
                \int\limits_{0}^{\infty}\int\limits_{W\pa}^{\infty}(w\pa-W\pa)^{3}
    \tilde{f}(w\pa,w\se)w\se\mathrm{d}w\pa\mathrm{d}w\se\mathrm{d}\vartheta
\end{align}
In the second integral we change the lower boundary $W\pa$ to zero and
subtract the integral from zero to $W\pa$, then we get finally
\begin{align}\label{eq:mphwp}
   q_{p,\Pi\pa}(W\pa) =2 n N \int\limits_{0}^{2\pi}
                \int\limits_{0}^{\infty}\int\limits_{0}^{W\pa}(W\pa-w\pa)^{3}
    \tilde{f}(w\pa,w\se)w\se\mathrm{d}w\pa\mathrm{d}w\se\mathrm{d}\vartheta
  + q_{p,\Pi\pa}(0) + 3 \Theta\pa P_{22p,\Pi\pa}(0)
\end{align}


\subsection{With: $W \pa=0,\vv{W} \se \ne \vv{0}$}
We assume first that $\vv{W}\pa=\mathrm{const}$.
For the integrals, containing $|\vv{w}+\vv{W}|$
or $|\vv{w}\se+\vv{W}\se|^{2n+1}$,
no analytic solution could be found, but in the anisotropic case we
have the form
$|\vv{w}\se+\vv{W}\se|^{2n}=(\vv{w}\se+\vv{W}\se)^{2n}$, with $\vv{W}\se=W_{1\se}\vv{e}_{1\se}+W_{2\se}\vv{e}_{2\se}$. We need
the integrals containing $n=1$ and $n=2$, for $n=1$ we find
\begin{align}\label{s1}
  (\vv{w}\se+\vv{W}\se)^{2} &= w\se^{2}+2
                                \vv{w}\se\vv{W\se}+W\se^{2}\\
  &= w\se^{2}+2 (w_{\se}W_{1\se}\cost+w_{\se}W_{2\se}\sint) +W\se^{2}
\end{align}
The integration over a full period of $\cost,\sint$ vanishes and we are
left only with the squares $w\se^{2},W\se^{2}$. Similar for $n=2$
\begin{align}\label{q1}
   (\vv{w}\se+\vv{W}\se)^{4} &= w\se^{4} + 4
         w\se^{2}\vv{w}\se\vv{W\se} + 4 W\se^{2}\vv{w}\se\vv{W\se}
         + 2 w\se^{2} W\se^{2} + W\se^{4}  + 4 (\vv{w}\se\vv{W\se})^{2}\\
  &=  w\se^{4}  + W\se^{4} +  2 w\se^{2} W\se^{2} + 4 (w_{\se}^{2}W_{1\se}^{2}\cost[2]+w_{\se}^{2}W_{2\se}^{2}\sint[2])
\end{align}
where we have dropped in the last line all terms including
$\cost,\sint$ or $\cost \sint$. 

Thus, the most probable speed and heat flow are 
given by:
\begin{align}
  u_{p\se}(W\se) &= n \Tp\Ts^{3} N \int\limits_{0}^{2\pi}
                \int\limits_{-\infty}^{\infty}\int\limits_{0}^{\infty}(w\se^{2}+W\se^{2})
                 \tilde{f}(w\pa,w\se)w\se\mathrm{d}w\pa\mathrm{d}w\se\mathrm{d}\vartheta\\
                     &= u_{p,\Pi\se} (\vv{0})+  n \Tp\Ts^{3} N
                       \mathcal{I}_{0} W\se\\
  q_{p\se} &=n \Tp\Ts^{5}N \int\limits_{0}^{2\pi}
                 \int\limits_{-\infty}^{\infty}\int\limits_{0}^{\infty}
                 (w\se^{4}  + W\se^{4} +  2 w\se^{2} W\se^{2} + 4
                 (w_{\se}^{2}W_{1\se}^{2}\cost[2]+w_{\se}^{2}W_{2\se}^{2}\sint[2])\\\nonumber
                     &\hspace{2cm}\tilde{f}(w\pa,w\se)\mathrm{d}w\se\mathrm{d}w\pa\mathrm{d}\vartheta\\
  &= q_{p\se} (\vv{0}) + 2\pi\Theta^{2}\se[1  + W_{\se}^{2}]
    u_{p\se}(\vv{0})+ n \Tp\Ts^{5}N \mathcal{I}_{0} W\se^{4} 
\end{align}
with
\begin{align}
  \mathcal{I}_{0} = \icyl f(w\pa,w\se)\mathrm{d}w\se\mathrm{d}w\se\mathrm{d}\vartheta
\end{align}

If we have the special case where $\vv{W}_{w\se} \parallel \vv{w}\se$ or
$\vv{W}_{w\se} = a \frac{\vv{w}\se}{w\se}$ with $a\in\mathbb{R}$, we can use
Eq.~\ref{s1} and~\ref{q1} to evaluate the most probable speed and heat
flow. Thus, we have:
\begin{align}
   u_{p\se}(\vv{W}_{w\se}) &= u_{p,\Pi\se}(\vv{0}) + a \Theta\se
                                  + a^{2}  n \Tp\Ts^{3} N
                       \mathcal{I}_{0} \\
    q_{p\se}(\vv{W}_{w\se}) &= q_{p,\Pi\se}(\vv{0}) + a\Theta\se
                                   P_{22\pi,se}(\vv{0}) + a^{2}\Theta\se^{2}u_{p,\Pi\se}(\vv{0})
                                   + a^{3}n \Theta\se^{3} +  a^{4}  n \Tp\Ts^{5}N \mathcal{I}_{0}  
\end{align}
The values for $\mathcal{I}_{0}$ are given in the table~\ref{tab:05}
in the main text. 


\begin{thebibliography}{}
\expandafter\ifx\csname natexlab\endcsname\relax\def\natexlab#1{#1}\fi
\providecommand{\url}[1]{\href{#1}{#1}}
\providecommand{\dodoi}[1]{doi:~\href{http://doi.org/#1}{\nolinkurl{#1}}}
\providecommand{\doeprint}[1]{\href{http://ascl.net/#1}{\nolinkurl{http://ascl.net/#1}}}
\providecommand{\doarXiv}[1]{\href{https://arxiv.org/abs/#1}{\nolinkurl{https://arxiv.org/abs/#1}}}

\bibitem[{{Abramowitz} \& {Stegun}(1972)}]{Abramowitz-Stegun-1972}
{Abramowitz}, M., \& {Stegun}, I.~A. 1972, {Handbook of Mathematical Functions}

\bibitem[{{Astfalk} \& {Jenko}(2016)}]{Astfalk-Jenko-2016}
{Astfalk}, P., \& {Jenko}, F. 2016, Journal of Geophysical Research (Space
  Physics), 121, 2842, \dodoi{10.1002/2015JA022267}

\bibitem[{{dos Santos} {et~al.}(2017){dos Santos}, {Ziebell}, \&
  {Gaelzer}}]{dosSantos-etal-2017}
{dos Santos}, M.~S., {Ziebell}, L.~F., \& {Gaelzer}, R. 2017, \apss, 362, 18,
  \dodoi{10.1007/s10509-016-2997-4}

\bibitem[{{Effenberger} {et~al.}(2012{\natexlab{a}}){Effenberger}, {Fichtner},
  {Scherer}, {Barra}, {Kleimann}, \& {Strauss}}]{Effenberger-etal-2012}
{Effenberger}, F., {Fichtner}, H., {Scherer}, K., {et~al.} 2012{\natexlab{a}},
  \apj, 750, 108, \dodoi{10.1088/0004-637X/750/2/108}

\bibitem[{{Effenberger} {et~al.}(2012{\natexlab{b}}){Effenberger}, {Fichtner},
  {Scherer}, \& {B{\"u}sching}}]{Effenberger-etal-2012b}
{Effenberger}, F., {Fichtner}, H., {Scherer}, K., \& {B{\"u}sching}, I.
  2012{\natexlab{b}}, \aap, 547, A120, \dodoi{10.1051/0004-6361/201220203}

\bibitem[{{Eliasson} \& {Lazar}(2015)}]{Eliasson-Lazar-2015}
{Eliasson}, B., \& {Lazar}, M. 2015, Physics of Plasmas, 22, 062109,
  \dodoi{10.1063/1.4922479}

\bibitem[{{Gradshteyn} \& {Ryzhik}(2007)}]{Gradstein-Ryshik-1981}
{Gradshteyn}, I.~S., \& {Ryzhik}, I.~M. 2007, Table of integrals, series, and
  products (Elsevier/Academic Press, Amsterdam), xlviii+1171

\bibitem[{{Kim} {et~al.}(2017){Kim}, {Schlickeiser}, {Yoon}, {L{\'o}pez}, \&
  {Lazar}}]{Kim-etal-2017}
{Kim}, S., {Schlickeiser}, R., {Yoon}, P.~H., {L{\'o}pez}, R.~A., \& {Lazar},
  M. 2017, Plasma Physics and Controlled Fusion, 59, 125003,
  \dodoi{10.1088/1361-6587/aa8898}

\bibitem[{{Lazar} {et~al.}(2016){Lazar}, {Fichtner}, \&
  {Yoon}}]{Lazar-etal-2016}
{Lazar}, M., {Fichtner}, H., \& {Yoon}, P.~H. 2016, \aap, 589, A39,
  \dodoi{10.1051/0004-6361/201527593}

\bibitem[{{Lazar} {et~al.}(2012){Lazar}, {Pierrard}, {Poedts}, \&
  {Schlickeiser}}]{Lazar-etal-2012}
{Lazar}, M., {Pierrard}, V., {Poedts}, S., \& {Schlickeiser}, R. 2012,
  Astrophysics and Space Science Proceedings, 33, 97,
  \dodoi{10.1007/978-3-642-30442-2_12}

\bibitem[{{Lazar} \& {Poedts}(2014)}]{Lazar-Poedts-2014}
{Lazar}, M., \& {Poedts}, S. 2014, \mnras, 437, 641,
  \dodoi{10.1093/mnras/stt1914}

\bibitem[{{Lazar} {et~al.}(2015){Lazar}, {Poedts}, \&
  {Fichtner}}]{Lazar-etal-2015}
{Lazar}, M., {Poedts}, S., \& {Fichtner}, H. 2015, \aap, 582, A124,
  \dodoi{10.1051/0004-6361/201526509}

\bibitem[{{Lazar} {et~al.}(2011){Lazar}, {Poedts}, \&
  {Schlickeiser}}]{Lazar-etal-2011}
{Lazar}, M., {Poedts}, S., \& {Schlickeiser}, R. 2011, \mnras, 410, 663,
  \dodoi{10.1111/j.1365-2966.2010.17472.x}

\bibitem[{{Lazar} {et~al.}(2017){Lazar}, {Shaaban}, {Poedts}, \& {{\v
  S}tver{\'a}k}}]{Lazar-etal-2017a}
{Lazar}, M., {Shaaban}, S.~M., {Poedts}, S., \& {{\v S}tver{\'a}k}, {\v S}.
  2017, \mnras, 464, 564, \dodoi{10.1093/mnras/stw2336}

\bibitem[{{Lazar} {et~al.}(2018){Lazar}, {Yoon}, {L{\'o}pez}, \&
  {Moya}}]{Lazar-etal-2018}
{Lazar}, M., {Yoon}, P.~H., {L{\'o}pez}, R.~A., \& {Moya}, P.~S. 2018, Journal
  of Geophysical Research (Space Physics), 123, 6, \dodoi{10.1002/2017JA024759}

\bibitem[{{Leubner} \& {Schupfer}(2002)}]{Leubner-Schupfer-2002}
{Leubner}, M.~P., \& {Schupfer}, N. 2002, Nonlinear Processes in Geophysics, 9,
  75

\bibitem[{{Maksimovic} {et~al.}(2005){Maksimovic}, {Zouganelis}, {Chaufray},
  {Issautier}, {Scime}, {Littleton}, {Marsch}, {McComas}, {Salem}, {Lin}, \&
  {Elliott}}]{Maksimovic-etal-2005}
{Maksimovic}, M., {Zouganelis}, I., {Chaufray}, J.-Y., {et~al.} 2005, Journal
  of Geophysical Research (Space Physics), 110, A09104,
  \dodoi{10.1029/2005JA011119}

\bibitem[{{Olbert}(1968)}]{Olbert-1968}
{Olbert}, S. 1968, in Astrophysics and Space Science Library, Vol.~10, Physics
  of the Magnetosphere, ed. R.~D.~L. {Carovillano} \& J.~F. {McClay}, 641

\bibitem[{Oldham {et~al.}(2010)Oldham, Myland, \& Spanier}]{Oldham-etal-2010}
Oldham, K., Myland, J., \& Spanier, J. 2010, An Atlas of Functions: with
  Equator, the Atlas Function Calculator, An Atlas of Functions (Springer New
  York).
\newblock \url{https://books.google.de/books?id=UrSnNeJW10YC}

\bibitem[{{Pierrard} \& {Lazar}(2010)}]{Pierrard-Lazar-2010}
{Pierrard}, V., \& {Lazar}, M. 2010, Sol.\ Phys., 267, 153,
  \dodoi{10.1007/s11207-010-9640-2}

\bibitem[{{Scherer} {et~al.}(2017){Scherer}, {Fichtner}, \&
  {Lazar}}]{Scherer-etal-2017}
{Scherer}, K., {Fichtner}, H., \& {Lazar}, M. 2017, EPL (Europhysics Letters),
  120, 50002, \dodoi{10.1209/0295-5075/120/50002}

\bibitem[{{Shaaban} {et~al.}(2018){Shaaban}, {Lazar}, {Astfalk}, \&
  {Poedts}}]{Shaaban-etal-2018}
{Shaaban}, S.~M., {Lazar}, M., {Astfalk}, P., \& {Poedts}, S. 2018, Journal of
  Geophysical Research (Space Physics), 123, 1754, \dodoi{10.1002/2017JA025066}

\bibitem[{{Shaaban} {et~al.}(2019){Shaaban}, {Lazar}, {L{\'o}pez}, {Fichtner},
  \& {Poedts}}]{Shaaban-etal-2019}
{Shaaban}, S.~M., {Lazar}, M., {L{\'o}pez}, R.~A., {Fichtner}, H., \& {Poedts},
  S. 2019, \mnras, 483, 5642, \dodoi{10.1093/mnras/sty3377}

\bibitem[{{Summers} \& {Thorne}(1991)}]{Summers-Thorne-1991}
{Summers}, D., \& {Thorne}, R.~M. 1991, Physics of Fluids B, 3, 1835,
  \dodoi{10.1063/1.859653}

\bibitem[{{{\v S}tver{\'a}K} {et~al.}(2008){{\v S}tver{\'a}K},
  {Tr{\'a}vn{\'{\i}}{\v c}ek}, {Maksimovic}, {Marsch}, {Fazakerley}, \&
  {Scime}}]{Stverak-etal-2008}
{{\v S}tver{\'a}K}, {\v S}., {Tr{\'a}vn{\'{\i}}{\v c}ek}, P., {Maksimovic}, M.,
  {et~al.} 2008, Journal of Geophysical Research (Space Physics), 113, A03103,
  \dodoi{10.1029/2007JA012733}

\bibitem[{{Vasyliunas}(1968)}]{Vasyliunas-1968}
{Vasyliunas}, V.~M. 1968, in Astrophysics and Space Science Library, Vol.~10,
  Physics of the Magnetosphere, ed. R.~D.~L. {Carovillano} \& J.~F. {McClay},
  622

\bibitem[{{Vi{\~n}as} {et~al.}(2015){Vi{\~n}as}, {Moya}, {Navarro}, {Valdivia},
  {Araneda}, \& {Mu{\~n}oz}}]{Vinas-etal-2015}
{Vi{\~n}as}, A.~F., {Moya}, P.~S., {Navarro}, R.~E., {et~al.} 2015, Journal of
  Geophysical Research (Space Physics), 120, 3307, \dodoi{10.1002/2014JA020554}

\bibitem[{{Yoon} {et~al.}(2018){Yoon}, {Lazar}, {Scherer}, {Fichtner}, \&
  {Schlickeiser}}]{Yoon-etal-2018}
{Yoon}, P.~H., {Lazar}, M., {Scherer}, K., {Fichtner}, H., \& {Schlickeiser},
  R. 2018, \apj, 868, 131, \dodoi{10.3847/1538-4357/aaeb94}

\bibitem[{{Ziebell} \& {Gaelzer}(2017)}]{Ziebell-Gaelzer-2017}
{Ziebell}, L.~F., \& {Gaelzer}, R. 2017, Physics of Plasmas, 24, 102108,
  \dodoi{10.1063/1.5002136}

\end{thebibliography}
\end{document}